%Paper: hep-th/9407025
%From: vafa@string.harvard.edu (Cumrun Vafa)
%Date: Tue, 5 Jul 1994 09:18:55 -0400

\input harvmac.tex

\def\frac#1#2{{#1\over#2}}

\mathchardef\ka="101A

\catcode`\@=11
\def\slash#1{\mathord{\mathpalette\c@ncel{#1}}}
\overfullrule=0pt

\def\steepslash{\c@ncel}
\def\frac#1#2{{#1\over #2}}

\def\MM{{\cal M}}

\def\XX{{\cal X}}

\def\inbar{\,\vrule height1.5ex width.4pt depth0pt}
\def\IB{\relax{\rm I\kern-.18em B}}
\def\IC{\relax\hbox{$\inbar\kern-.3em{\rm C}$}}
\def\IP{\relax{\rm I\kern-.18em P}}
\def\IR{\relax{\rm I\kern-.18em R}}
\def\IZ{\relax\ifmmode\mathchoice
{\hbox{Z\kern-.4em Z}}{\hbox{Z\kern-.4em Z}}
{\lower.9pt\hbox{Z\kern-.4em Z}}
{\lower1.2pt\hbox{Z\kern-.4em Z}}\else{Z\kern-.4em Z}\fi}

\catcode`\@=12

\Title{\vbox{\hbox{HUTP--94/A016,}}
       \vbox{\hbox{IASSNS-HEP-94/47}}   }
{\vbox{\centerline{\bf Superstrings and Manifolds of Exceptional Holonomy}}}
\bigskip\centerline{{ Samson L. Shatashvili} \footnote{$^\#$}
{On leave of absence from St. Petersburg
Branch of Mathematical Institute (LOMI), Fontanka 27, St.Petersburg
191011, Russia.}}
\bigskip\centerline{School of Natural Sciences,
Institute for Advanced Study}
\centerline{Olden Lane, Princeton, NJ 08540}
\bigskip\centerline{ Cumrun Vafa }
\bigskip\centerline{Lyman Laboratory of Physics, Harvard University}
\centerline{Cambridge, MA 02138, USA}
\centerline{and}
\centerline{School of Natural Sciences,
Institute for Advanced Study}
\centerline{Olden Lane, Princeton, NJ 08540}
 \vskip .1in
The condition of having an $N=1$ spacetime supersymmetry
for heterotic string leads to 4 distinct possibilities
for compactifications namely compactifications down to 6,4,3 and 2 dimensions.
Compactifications to 6 and 4 dimensions have been studied
extensively before (corresponding to $K3$ and a Calabi-Yau
threefold respectively).  Here we complete the
study of the other two cases corresponding to compactification
down to 3 on a 7 dimensional manifold of $G_2$ holonomy and compactification
down to 2 on an 8 dimensional manifold of $Spin(7)$ holonomy.  We study
the extended chiral algebra and find the space of exactly marginal
deformations.  It turns out that the role the $U(1)$ current plays in the
$N=2$ superconformal theories, is played by tri-critical Ising model
in the case of $G_2$ and Ising model in the case of $Spin(7)$ manifolds.
  Certain generalizations of mirror symmetry
are found for these two cases.  We also discuss the topological
twisting in each case.

%\draft
\Date{July, 94}

\newsec{Introduction}
Supersymmetric sigma models in two dimensions have been the source of
many interesting ideas in the interplay between quantum
field theories and geometry and topology of manifolds.
  In the context
of superstring theories, viewing strings moving on a manifold
leads to the use of sigma models as the building blocks of string
vacua.  To be a string vacuum the sigma model must lead
to a conformal theory in two dimensions.  Moreover to lead
to spacetime supersymmetry,
 which is the only class of superstrings
we know which are perturbatively stable, the manifold should
admit covariantly constant spinors which can be used to define
the supersymmetry transformation.  It turns out that having
a covariantly constant spinor already guarantees conformal
invariance to one loop order in the
sigma model perturbation theory and perhaps
to all orders (with appropriate adjustments of the metric),
 so the study of manifolds admitting covariantly constant
spinors seems like an important question for string theory.
In general if we have an $n$ dimensional Riemannian manifold the
holonomy group is in $SO(n)$; however having a covariantly
constant spinor the holonomy group is smaller and it is (a subgroup of)
the little group which leaves a spinor of $SO(n)$ invariant.

Since superstrings live in 10 dimensions and we in 4, the
most important physical case to study is the 6 dimensional
manifolds with covariantly constant spinors.  If we require
only one spacetime supersymmetry, this means we need only
one covariantly constant spinor (for a fixed chirality) and this
leads to the manifolds of $SU(3)$ holonomy, i.e. the Calabi-Yau
manifolds \ref\chsw{P. Candelas, G. Horowitz, A. Strominger and E. Witten,
Nucl. Phys., B258, (1985) 46.}. These manifolds have
been investigated a great deal with interesting physical results.
Among these one could mention that many
classically singular Calabi-Yau manifolds lead to non-singular
sigma models. Also there is a mirror phenomenon which means
that strings on two inequivalent manifolds can lead to the same
sigma model.
Moreover there is a topological ring in these theories
known as the chiral ring \ref\lvw{W. Lerche, C. Vafa and N. Warner,
Nucl. Phys., B324 (1989) 427.}\
 which captures the deformation structure of the Calabi-Yau
manifold.

In fact the sigma models based on
 Calabi-Yau manifolds have been studied  for all dimensions and not
just 6, and many important aspects of the theory behave
uniformly well in all dimensions.
Also, one could consider odd dimensional manifolds with a minimal
number of covariantly constant spinors
by considering
the product of Calabi-Yau with a circle.
However, if one is interested in the minimum number
of supersymmetries allowed
this class misses two special cases (for a review see
\ref\mckenz{McKenzie Y. Wang, Ann. Global Anal. Geom. 7 (1989) 59})
:  First of all, in manifolds
of 7 dimensions the minimum number of supersymmetries is
given by a manifold of $G_2$ holonomy which has only one covariantly
constant spinor as opposed to
a manifold of $SU(3)$ holonomy times a circle which has 2.
Furthermore in dimension 8 an $SU(4)$ holonomy manifold
leads to 2 covariantly constant spinors whereas
for an 8 dimensional manifold of $Spin(7)$ holonomy there
will be only 1 covariantly constant spinor.  The
possible existence of these
 two special
cases had been known for a long time \ref\ber{M.
Berger, Bull. Soc. Math. France 83 (1955) 279}.  It is
very amusing that these
two special cases can be used in physical models
simply because the dimensions where they occur is less than 10,
 which means that if we were to compactify
superstrings down to 3 or 2 dimensional Minkowski space and
ask which manifolds would lead to minimal number of nonvanishing
supersymmetries
(1 for heterotic strings and 2 for type II strings) we would have
to study sigma models on 7d manifolds of $G_2$ holonomy and 8d manifolds of
$Spin(7)$ holonomy.  The study of these two classes
of sigma models is the subject of the present paper.

The organization of this paper is as follows:  In section
2 we will review basic facts about superconformal theories
in general and manifolds with covariantly constant spinors
in particular.  This includes a quick review of aspects
of N=2 superconformal theories and their relation to geometry
of Calabi-Yau manifolds.  This review is a good exercise
for setting the stage for the two special cases of manifolds
of $G_2$ and $Spin(7)$ holonomy and the associated conformal theories.
 In this section we also review some geometrical facts about
manifolds of $G_2$ and $Spin(7)$ holonomy that we will need
in the rest of the paper.

In section 3
using the geometrical data at hand
we construct
the extended chiral algebra associated to $G_2$ and $Spin(7)$ manifolds.
It turns out that the role played by the $U(1)$ piece of the
$N=2$ algebra in the context of Calabi-Yau compactification is
now played by the tri-critical Ising model for $G_2$ manifolds
and Ising model for $Spin(7)$ manifolds!
The algebra and its construction is very similar in both cases
and will be discussed in parallel.  The existence of these
two minimal models as an integral part of the theory
is crucial. In particular it allows
us to identify the space of marginal operators
which preserve both the superconformal symmetry
as well as the $G_2$ and $Spin(7)$ structure of the algebra
and prove their exact marginality to all orders in conformal
 perturbation theory.  Moreover we find the identification
of these deformations with the geometrical facts explained in section 2.

In section 4 we discuss concrete orbifold examples of these manifolds
constructed very recently by Joyce.
These examples are illuminating as far as the structure of the
algebra we found in the previous section.  Moreover we find a special
kind of mirror phenomenon takes place, in that we find inequivalent
orbifold resolutions
(having different betti numbers) found
 by Joyce correspond to the same underlying
conformal theory up to deformation in moduli space.
  Moreover we find that whenever there
are discrete torsions which lead to different
conformal theories there are also inequivalent geometrical resolutions.

In section 5 we construct a topological twisting for these cases.
Again amazing facts
about tri-critical Ising model and Ising model are crucial
for making this twist possible.
In appendix 1 and appendix 2 we collect some relevant facts about the
structure of the extended chiral algebras that we have
encountered for these exceptional manifolds.

\newsec{Superconformal Sigma Models and Special Holonomy Manifolds}

In this section we review some general aspects of 2d supersymmetric sigma
models and their interplay with geometry of the target manifolds.
The most basic observation in this regard is that
if we consider the Hilbert space on a circle
with periodic boundary conditions (the Ramond sector) of a 2d supersymmetric
sigma model with an $n$-dimensional target space $M$,
or for that matter the 1d supersymmetric sigma model on $M$,
there is an identity
for Witten's index \ref\witeul{E. Witten, Nucl. Phys., B202 (1982) 253.}
$${\rm Tr}(-1)^F {\rm exp}(-\beta {\rm H})=\chi (M) =\sum_{i=0}^{n}
(-1)^ib_i=n_+-n_-,$$
where $\chi(M)$ is the Euler characteristic of $M$
and $b_i$ are the betti numbers of $M$
and $n_+$ and $n_-$ denote the total
number of even and odd dimensional cohomologies respectively.  The basic idea
is that only the ground states with $H=0$
contribute to the index (as the $H>0$ come
in pairs with opposite $(-1)^F$)
and that in a suitable limit
the ground states are related to the harmonic forms on $M$, and
$(-1)^F$, up to an overall sign ambiguity, can be
identified with the parity of the degree of the harmonic forms. Actually there
is more information \ref\witmor{E. Witten, J. Diff.
Geometry, 17 (1982) 661}:
It is possible to show that the number of ground states in the theory
are exactly equal to the number of harmonic forms. In other words
all the non-cohomological perturbative ground states are lifted
up by non-perturbative effects but the other ground states, which
are in equal number to the cohomology elements are
exact non-perturbative ground states and are not paired to become
massive even if they have opposite $(-1)^F$, i.e.,
$${\rm Tr}\ {\rm exp}(-\beta {\rm H})\bigg|_{\beta \rightarrow \infty}=
\sum_{i=0}^{n}b_i=n_++n_-.$$
Note that even though the number of ground states are equal to the
number of cohomology elements of $M$ there is no canonical correspondence.
In particular it is not in general possible to determine the betti numbers
individually.  From the two physical computations above
we can only deduce $n_+$ and $n_-$.  Actually even that
we can not deduce unambiguously because, as was mentioned before,
the sign of $(-1)^F$ cannot be canonically fixed.  Therefore
from the physical Hilbert space we can only deduce $n_+$ and $n_-$
up to the exchange $n_+\leftrightarrow n_-$.
So much is true for general supersymmetric sigma models.  If
there are further restrictions on $M$ we can deduce more from
the physical theory.  For example if $M$ is a
K\"ahler manifold, the fermions are complex and so
there is a $U(1)$ conserved charge corresponding
to the fermion number $F$ which acting on the ground state
can be identified with the number of holomorphic forms $p$
minus the number of antiholomorphic forms $q$ of the harmonic
form
$$F=p-q.$$
  So in this way by decomposing the ground state to eigenstates
of $F$ we can compute the number of cohomology elements with a given
value of $p-q$. There is also the chiral (or axial) fermion number
$F_A$ which
is non-perturbatively conserved only when the first chern class
$c_1(M)=0$\foot{For the 1d supersymmetric sigma models since
there are no instantons to ruin the conservation,
the $F_A$ is always conserved.}, i.e.,
when $M$ is a Calabi-Yau manifold. $F_A$ can be identified
when acting on ground states with
$$F_A=p+q-d,$$
where $d=n/2$ is the complex dimension of $M$.
So we can compute
$$p-{d\over 2}={1\over 2}(F_A+F)=F_L,$$
$$q-{d\over 2}={1\over 2}(F_A-F)=F_R.$$
Just as before
there is a relative ambiguity in the identification
of the sign of $F_{L,R}$ which
means that we can determine the hodge numbers
$h^{p,q}$ only up to the ambiguity
$$h^{p,q}\leftrightarrow h^{d-p,q}.$$
This apparent deficiency in the supersymmetric
sigma model in capturing geometry of Calabi-Yau was
 conjectured \lvw \ref\dix{L. Dixon, {\it Some world sheet properties of
superstring compactifications, on orbifolds and otherwise},
lecture given at the 1987 ICTP Summer Workshop, Trieste, Italy, 1987.}\
to be related to the beautiful possibility
that CY manifolds may come in pairs which lead to the
same sigma model but for which the hodge numbers are
mirror to each other.  There is by now a large body
of evidence supporting this conjecture \ref\mirrorb{Essays on Mirror Manifolds,
editor S.-T. Yau, International Press, 1992.}.

There are more interesting relations with geometry of CY if we
consider operator products of some special operators in the theory.
It turns out that there is a natural ring of operators in the theory
known as the {\it chiral ring} \lvw\
which are in one to one correspondence with the cohomology elements and
which are most easily defined by using the fact that there exists
a metric on CY which
gives rise to a 2d conformal field theory.  One can also define
this ring by purely topological means \ref\wittop{
E. Witten, Nucl. Phys. B340 (1990) 281.}, and it turns
out that (at least in one version) it is related to a
{\it quantum deformed} cohomology ring of the manifold
which has the information about the holomorphic maps
from $CP^1$ to $M$ encoded in it.

Calabi-Yau 3-folds are interesting for string
theory as mentioned in the introduction precisely because
they have covariantly constant spinors and they
have a minimal number of them leading
to $N=1$ spacetime supersymmetry when we
compactify heterotic strings on them.  However let us ask
a question which would be a very natural question in the
context of superstring or heterotic string compactifications:
If we compactify the heterotic strings to any lower dimension,
in which dimensions can we obtain the minimal non-vanishing number of
 spacetime supersymmetries?
The answer to this question is rather interesting
(for a simple derivation see \mckenz ):
To have one spacetime supersymmetry
we need the minimum allowed covariantly constant
spinors (1 or 2 depending on the dimension).  This is possible only
if we compactify from 10 down to 6,4,3 or 2 on manifolds
with holonomy $Sp(1)(=SU(2))$, $SU(3)$, $G_2$ and $Spin(7)$,
with dimensions 4,6,7,8 respectively.  Moreover in the case
of 4 dimensions there is a unique manifold $K3$ which
has $Sp(1)$ holonomy.  The six dimensional
case is a three fold Calabi-Yau and possibilities
for this has also been studied extensively.  Here we
begin the completion of this systematic classification
by studying sigma models on manifolds with $G_2$ and $Spin(7)$
holonomy. It is amusing to note that the $K3$ and
CY threefolds have generalization to higher dimensions with manifolds
of $Sp(n)$ and $SU(n)$ holonomy respectively, but the $G_2$ and
$Spin(7)$ case are unique structures with no generalizations
to higher dimensions.

Before we go on to describe some general properties of manifolds
with $G_2$ and $Spin(7)$ holonomy,
motivated by the success of the mirror conjecture for CY target
spaces let us make a conjecture
which will prove helpful in clarifying the observations we shall
make later:

{\it Generalized Mirror Conjecture}: The degree of ambiguity
left by being unable to decipher all the topological
aspects of the target manifold using the algebraic formulation
of quantum field theories is precisely explained by having
topologically inequivalent manifolds allowed by the ambiguity
to lead to the same quantum field theory up to deformation
in the moduli of the quantum field theory.

We shall see in later sections
the first non-Calabi-Yau examples which support
the above conjecture in the case of manifolds of $G_2$
and $Spin (7)$ holonomy.

Let us begin discussing some facts about manifolds of $G_2$ and
$Spin(7)$ holonomy.  Until very recently the only known examples
of manifolds of $G_2$ holonomy and $Spin(7)$ holonomy were
non-compact manifolds \ref\br{R. L. Bryant, Ann. Math. 126 (1987)
525\semi R. L. Bryant and S. M. Salamon,
Duke Math. J. 58 (1989) 829.}.  The situation
has dramatically changed recently due to the work of Joyce
\ref\joyce{D. D. Joyce, {\it Compact 7-manifolds with holonomy}
$G_2$,I,II, IAS preprints 1994; {\it Compact
Riemannian 8-manifolds with Exceptional Holonomy} $Spin(7)$,
in preparation.}\ who constructed the first compact
examples of manifolds with $G_2$ and $Spin(7)$ holonomy,
which we denote by $M^7$ and $M^8$ respectively.
Just as in the Calabi-Yau case where the fact that the
manifold has $SU(n)$ holonomy leads to the existence of a unique
non-vanishing holomorphic covariantly constant $n$-form
(and of course its conjugate), in these two exceptional
cases a similar thing happens (see \ref\sal{S. L. Salamon,
{\it Riemannian geometry and holonomy groups}, Pitman
Research notes in mathematics series no. 201, published
by Longman, Harlow (1989)} Chapters 11 and 12):
In the case of $G_2$ manifolds,
there is a canonical 3 form $\phi$ and its dual which is a 4 form
$*\phi$ which are covariantly constant and
in the case of $Spin(7)$ there is a self-dual 4-form
$\Omega$.  They can be locally written
as follows:
we choose a local vielbein so that the metric is
$\sum e_i\otimes e_i$ where $e_i$ are one forms
and for the $G_2$ case, $i$ runs from $1$ to $7$ and for
the $Spin(7)$ case $i$ runs from $1$ to $8$.   Then
these forms can be written as
\eqn\ffi{\phi =e_1\wedge e_2 \wedge e_7+e_1\wedge e_3 \wedge_6
+e_1\wedge e_4 \wedge e_5 +e_2\wedge e_3\wedge e_5 -e_2
\wedge e_4 \wedge e_6 +e_3\wedge e_4 \wedge e_7 +e_5 \wedge e_6\wedge e_7,}
\eqn\dfi{\eqalign{*\phi=e_1\wedge e_2\wedge e_3\wedge e_4+ &e_1 \wedge e_2
\wedge e_5 \wedge e_6 -e_1\wedge e_3\wedge e_5 \wedge e_7+
e_1 \wedge e_4 \wedge e_6 \wedge e_7 +\cr
&e_2 \wedge e_3 \wedge e_6 \wedge e_7
+e_2 \wedge e_4 \wedge e_5 \wedge e_7 +
e_3\wedge e_4\wedge e_5 \wedge e_6,}}
\eqn\ome{\Omega =e_8 \wedge {\phi} -*\phi.}
These can be understood as follows: In the case of $G_2$
if we view $e_i$ as forming
the fundamental representation of $O(7)$, the fact that the holonomy
is in the $G_2$ subgroup of $O(7)$ means that in the threefold
tensor product of this representation there is a totally anti-symmetric
singlet of $G_2$ which is identified with $\phi$. Similarly
$*\phi$ is invariant under $G_2$.
In the case of 8 dimensional
$Spin(7)$ manifolds $e_i$ form the fundamental representation
of $O(8)$.  If we view the embedding of $Spin(7)$ in $O(8)$ such
that the $8$ dimensional spinor representation of $O(8)$ transforms
as the $7\oplus 1$ of $Spin(7)$, and thus the $8$ dimensional
vector representation of $O(8)$ transforms as an eight dimensional
 spinor representation of $Spin(7)$, then
 in the fourfold totally antisymmetric product
of this latter representation there is a unique singlet of
$Spin(7)$ which is denoted by $\Omega$ above.
 Metric $\sum e_i\otimes e_i$ can be uniquely reconstructed from
$\phi$ and $\Omega$.

Moreover it is true \ref\unpub{R. L. Bryant and F. R. Harvey,
unpublished}\joyce\ that the dimension
of moduli space of deformation of manifolds of $G_2$ holonomy
is $b_3(M^7)$ and the dimension of the moduli space
of deformation of manifolds of $Spin(7)$ holonomy is $b_4^-(M^8)+1$,
where $b_4^{\pm}$ denote the self-dual/anti-self-dual dimensions
of $H^4(M^8)$.  The simplest class of examples considered
by Joyce involve toroidal orbifolds.  In the case of $G_2$,
the minimal example is obtained by modding out $T^7/(Z_2)^3$
where each $Z_2$ has for eigenvalues of holonomy $(-1,-1,-1,-1,1,1,1)$,
but they sit in $SO(7)$ in such a way that they cannot be embedded
in an $SU(3)$ subgroup of it, but can be embedded into a $G_2$
subgroup of it.  Moreover it is clear from the above discussion
that this group will preserve $\phi$ and $*\phi$.  Moreover
for simplicity of analysis, Joyce considers some
of these $Z_2$'s to be accompanied with translations of the $T^7$,
and shows that the singular orbifold can be desingularized.
For the case of $Spin(7)$ holonomy the simplest examples he constructs
involve again desingularizing a toroidal orbifold. In this
case he considers $T^8/(Z_2)^4$, where each $Z_2$ has eigenvalues
of holonomy $(-1,-1,-1,-1,1,1,1,1)$, but again in such a way
that the full group does not sit in $SU(4)$ but does sit
in a $Spin(7)$ subgroup of $O(8)$.  In both the $G_2$
case and the $Spin(7)$ case he finds that there
are in general many inequivalent ways of desingularizing the manifold,
which we will be able to explain physically in section 4
as a consequence of the generalized mirror conjecture stated above.
In fact it is crucial to note that the dimension of the moduli
space of the conformal theory is actually {\it bigger} than
that predicted geometrically.  The reason for this is that
the possibility of using the anti-symmetric two form to add
a phase to the action has no geometrical analog.  Therefore we have
$${\rm dim.\  moduli}_{physical}={\rm dim.\  moduli}_{geometrical} +b_2.$$
In particular for the $G_2$ case the dimension of sigma model
moduli is $b_2+b_3$ and not $b_3$ and for the case of $Spin(7)$ the
dimension of sigma model moduli is $b_4^-+b_2+1$.
Let us also briefly talk about the structure of the betti
numbers of these two cases:  In both cases we are dealing
with manifolds with $b_1=0$ in order to obtain the minimum
number of covariantly constant spinors.  In the case of
$G_2$ holonomy, therefore there are two independent betti
numbers to compute $b_2$ and $b_4$, since by duality $b_3=b_4$
and $b_5=b_2$.  As discussed before physically we can a priori
only compute the dimension of even or odd cohomologies
$b_2+b_4=b_5+b_3$. So a priori physically
we can expect to deduce only one geometrical index in this
case namely $b_2+b_4$ which is also equal to $b_2+b_3$ which
is the dimension of moduli space.

 In the case of $Spin(7)$ holonomy manifolds
there are a priori four topological numbers one can
hope to compute $b_2,b_3,b_4^{\pm}$
and the rest are obtained by duality.  However the fact that
there is a unique zero mode for the Dirac operator implies
using the index theorem that \ref\wangi{M. Y. Wang, Ind. Univ. Math.
J. 40 (1991) 815.}\
\eqn\rest{b_3+b_4^+-b_2-2b_4^--1=24.}
So geometrically there are only three independent numbers in this case.
Physically to begin with we have the number of even and odd cohomologies
which we can deduce $b_2+b_4+b_6=2b_2+b_4$ and $b_3+b_5=2b_3$
We should also in addition expect to compute $b_4^-+b_2+1$
by finding the dimension of exactly marginal deformations.
However the relation \rest\ implies that there
is a linear relation between these numbers which would mean that there
are only two independent physical numbers one could hope to compute,
as opposed to three in the geometrical case.
The validity of \rest\ for sigma model should follow from modular
invariance type arguments in relation to sigma models
\ref\shelwar{A. N. Schellekens and N. Warner, Nucl. Phys, B287 (1987) 317.}.

\newsec{Extended Symmetry Algebra, Consequences and Deformations}

In this section we will unravel the extended symmetry algebras
which underlie sigma models with $N=1$ superconformal
symmetry on manifolds of $G_2$ and $Spin(7)$ holonomy.
The idea for obtaining these symmetry algebras is familiar
from the study of Calabi-Yau manifolds, where one appends
to the $N=1$ superconformal algebra, a $U(1)$ current to obtain
the $N=2$ algebra, and the spectral flow operator, to guarantee
integrality of $U(1)$ charges. For the case of Calabi-Yau
three folds this has been studied in \ref\odake{S. Odake, Mod. Phys. Lett.,
A4 (1989) 557.}.
We study the
representation theory of this extended algebra. Consequences of this symmetry
allows us to gain insight into the structure of the theory and in particular
construct the space of exactly marginal deformations (the moduli).

Perhaps to make some aspects of the
 algebra that we obtain a little less
mysterious it would be helpful to see a priori what we should expect to play
the role that $U(1)$ plays for sigma models on manifolds of $SU(n)$
holonomy\foot{This line of thought was developed following a suggestion
of E. Martinec.}. If we start with a sigma model on a K\"ahler
manifold we have a priori a $U(n)$ symmetry.  Having a holonomy
in $SU(n)$ means that part of the symmetry is broken but
we are left with an unbroken $U(1)=U(n)/SU(n)$.  Similarly
in the case of 7 dimensional manifolds of $G_2$ holonomy,
a priori we have $SO(7)$ symmetry (more precisely $SO(7)$
current algebra at level 1).  The holonomy of the manifold
being in $G_2$ means that we are left with the residual
symmetry $SO(7)/G_2$, which in geometrical terms is no longer
a group, however, from the viewpoint of conformal theory it
is a coset model.  Computing its central charge we see that
since $SO(7)$ at level $1$ has central charge 7/2 and $G_2$
at level 1 has central charge $14/5$, the central charge of the
residual system is
$${7\over 2}-{14\over 5}={7\over 10},$$
which is thus a tri-critical Ising model
 \ref\go{P. Goddard and D. Olive,
Nucl. Phys., B257 (1985) 226.}!  Similarly for the
case of $Spin(7)$ manifolds one considers $SO(8)/Spin(7)$
which gives a central charge
$$4-7/2=1/2,$$
which is just the Ising model.  Below we shall recover these
facts directly as well as find out that these symmetries mix
in a very interesting way with the $N=1$ superconformal
algebra to obtain the extended symmetry algebra of our models.

\subsec{$G_2$}
As far as the algebraic structure is concerned we start from the
flat 7 dimensional space, and construct the chiral operators which
we expect to exist even after we perturb the metric to obtain
a non-trivial $G_2$ holonomy.  We of course expect to have
the energy momentum tensor $T$ and its superpartner $G$.  Moreover
the fact that a three form $\phi$ exists even after the perturbation
suggests that one can add to $N=1$
superconformal generators $T=T_b+T_f=
\frac{1}{2}{\sum}^7_1:J^1J^1:-\frac{1}{2}{\sum}_1^7:\psi^i{\partial}\psi^i$,
$G={\sum}_1^7:J^i\psi^i:$ a new spin 3/2 operator
\eqn\a{\eqalign{\Phi = &{\psi}^1{\psi}^2{\psi}^5 + {\psi}^1{\psi}^3{\psi}^6 +
{\psi}^1{\psi}^4{\psi}^7 - {\psi}^2{\psi}^3{\psi}^7 +
{\psi}^2{\psi}^4{\psi}^6 - {\psi}^3{\psi}^4{\psi}^5 +\cr
& {\psi}^5{\psi}^6{\psi}^7=f_{ijk}\psi^i\psi^j\psi^k,}}
with the coefficients $f_{ijk}$ defined by the
$G_2$ invariant three form $\phi$
from the previous section. We use the notation $J^i={\partial}x^i$ with
$x^i$ being a bosonic sigma model coordinate.
$N=1$ generators are invariant under the rotation
group $SO(7)$ and  $\Phi$ is invariant only under the $G_2$ subgroup of
$SO(7)$. If we compute the operator expansion of new generator
$\Phi$ with itself we obtain:

\eqn\ff{\Phi(z)\Phi(w) = -\frac{7}{(z-w)^3} + \frac{6}{z-w} X(w),}
where operator $X$ has spin 2

\eqn\c{\eqalign{X = &- {\psi}^1{\psi}^2{\psi}^3{\psi}^4 +
{\psi}^1{\psi}^2{\psi}^6{\psi}^7 -
{\psi}^1{\psi}^3{\psi}^5{\psi}^7 +
{\psi}^1{\psi}^4{\psi}^5{\psi}^6 - \cr
&{\psi}^2{\psi}^3{\psi}^5{\psi}^6 - {\psi}^2{\psi}^4{\psi}^5{\psi}^7 -
{\psi}^3{\psi}^4{\psi}^6{\psi}^7 - 1/2 :{\partial}{\psi}^i {\psi}^i:=
-*\Phi + T_f,}}
and is a linear combination of `dual' operator $*\Phi$ (defined by dual form
$*\phi$) and a fermionic stress-tensor.
Next step is to compute operator expansion
of the operators $X$ and $\Phi$. We obtain:

\eqn\fx{\Phi(z)X(w) = -\frac{15}{2}\frac{1}{(z-w)^2} \Phi(w) -
\frac{5}{2}\frac{1}{z-w} {\partial}\Phi(w),}

\eqn\xx{X(z)X(w) = \frac{35}{4}\frac{1}{(z-w)^4} - \frac{10}{(z-w)^2} X(w) -
\frac{5}{z-w} {\partial}X(w).}
This is not the end of
 story because now we need to deal with superpartners of
new generators with respect to original $N=1$ algebra. This introduces two new
operators of spins 2 and $\frac{5}{2}$ into the game;
we will denote them by $K$ and $M$ respectively:
\eqn\gf{G(z)\Phi(w) = \frac{1}{z-w}K(w),}
\eqn\gx{G(z)X(w) = -\frac{1}{2}\frac{1}{(z-w)^2} G(w) + \frac{1}{z-w} M(w).}
New operators in the right hand side have the following
free field representation:

\eqn\b{\eqalign{K = &J^1{\psi}^2{\psi}^5 + J^1{\psi}^3{\psi}^6 +
J^1{\psi}^4{\psi}^7 -
J^2{\psi}^1{\psi}^5 -
J^2{\psi}^3{\psi}^7 +\cr
& J^2{\psi}^4{\psi}^6 - J^3{\psi}^1{\psi}^6 + J^3{\psi}^2{\psi}^7 -
J^3{\psi}^4{\psi}^5 - J^4{\psi}^1{\psi}^7 - J^4{\psi}^2{\psi}^6 +\cr
& J^4{\psi}^3{\psi}^5 + J^5{\psi}^1{\psi}^2 - J^5{\psi}^3{\psi}^4 +
J^5{\psi}^6{\psi}^7 + J^6{\psi}^1{\psi}^3 +
J^6{\psi}^2{\psi}^4 -\cr
& J^6{\psi}^5{\psi}^7 + J^7{\psi}^1{\psi}^4 - J^7{\psi}^2{\psi}^3 +
J^7{\psi}^5{\psi}^6,}}

\eqn\d{\eqalign{M = &- J^1{\psi}^2{\psi}^3{\psi}^4 +
J^1{\psi}^2{\psi}^6{\psi}^7 - J^1{\psi}^3{\psi}^5{\psi}^7 +
J^1{\psi}^4{\psi}^5{\psi}^6 + J^2{\psi}^1{\psi}^3{\psi}^4 - \cr
&J^2{\psi}^1{\psi}^6{\psi}^7 - J^2{\psi}^3{\psi}^5{\psi}^6 -
J^2{\psi}^4{\psi}^5{\psi}^7 -J^3{\psi}^1{\psi}^2{\psi}^4 +
J^3{\psi}^1{\psi}^5{\psi}^7 +\cr
& J^3{\psi}^2{\psi}^5{\psi}^6 - J^3{\psi}^4{\psi}^6{\psi}^7 +
J^4{\psi}^1{\psi}^2{\psi}^3 - J^4{\psi}^1{\psi}^5{\psi}^6 +
J^4{\psi}^2{\psi}^5{\psi}^7 + \cr
&J^4{\psi}^3{\psi}^6{\psi}^7 - J^5{\psi}^1{\psi}^3{\psi}^7 +
J^5{\psi}^1{\psi}^4{\psi}^6 - J^5{\psi}^2{\psi}^3{\psi}^6 -
J^5{\psi}^2{\psi}^4{\psi}^7 +\cr
&  J^6{\psi}^1{\psi}^2{\psi}^7 - J^6{\psi}^1{\psi}^4{\psi}^5 +
J^6{\psi}^2{\psi}^3{\psi}^5 - J^6{\psi}^3{\psi}^4{\psi}^7 -
J^7{\psi}^1{\psi}^2{\psi}^6 +\cr
& J^7{\psi}^1{\psi}^3{\psi}^5 + J^7{\psi}^2{\psi}^4{\psi}^5 +
J^7{\psi}^3{\psi}^4{\psi}^6 + 1/2 J^i{\partial}{\psi}^i - 1/2
{\partial}J^i{\psi}^i .}}

A nontrivial fact deeply related to `$G_2$ structure' is that operator
expansion algebra formed by these six operators $T, G, \Phi, X, K$ and $M$
closes. (
The results of further computation is presented in the  Appendix 1 together
with commutation relations written in mode expansion.)
Thus, we have demonstrated that there is an extended chiral
algebra which contains quadratic combinations
in the right hand side and thus reminds (just reminds)
one of
$W$-algebra.\foot{The
existence of extended symmetry for $N=1$ sigma model on $G_2$ manifold
in classical approximation was previously mentioned  in \ref\howe{P. S. Howe
and G. Papadopolous, Comm. Math. Phys., 151 (1993) 467.}; we also
have been informed by M. Rocek and J. de Boer that recently they also have
 found extended symmetry in above sigma model.}

After extended chiral algebra is derived we can forget about the free
field picture recalling that the perturbation will destroy the fact
that the theory is free, but assume the existence of the algebra
beyond free realization
and study the corresponding conformal field theory. As a first step
we have to find the spectrum of low lying
states and in particular the spectra of Ramond
ground states which carry the geometrical information
about the manifold. In this study it is extremely
useful to note that our extended algebra contains two (non-commutative)
$N=1$ superconformal subalgebras:
1. Original $N=1$ generated by $G$ and $T$, and 2. $N=1$ superconformal
algebra generated by $G_I=\frac{i}{\sqrt 15}\Phi$ and $T_I=
-\frac{1}{5}X$.
Moreover, the latter is a very
interesting one - it has a Virasoro central charge $\frac{7}{10}$
as predicted in the beginning of this section and is
the tri-critical Ising model which is the only bosonic minimal model in the
list of $N=1$ superconformal minimal models \ref\fqs{D. Friedan, Z. Qiu and
S. Shenker, Phys. Lett., 151B, (1985) 37; M Bershadskii, V. Knizhnik and
M. Teitelman, Phys. Lett., 151B (1985) 31.}.
In addition a simple observation that
\eqn\simp{T_I(z)T_r(w) = 0, \quad \quad  T=T_I+T_r}
allows us to classify the highest weight representations of our algebra
using two numbers: Tri-critical Ising highest weight and eigenvalue of the
zero mode of the remaining stress-tensor $T_r$.

Now, at the beginning we consider only chiral sector (left-movers say).
The theory is supersymmetric and thus we have two sectors -
Neveu-Schwarz and Ramond. We shall see below that the $(-1)^F$ for the
full theory can be identified with the $(-1)^{F_I}$ which
is the $Z_2$ symmetry of the tri-critical Ising model viewed as
an $N=1$ superconformal system.
{}From the observation that total stress tensor
can be written as a sum of two commutative Virasoro generators where one is
tri-critical Ising, we conclude that unitary highest weight
representations should have following tri-critical Ising dimensions:

\eqn\nsr{NS: \quad \quad \quad \quad [0]_{Vir}, \quad \quad \quad
[\frac{1}{10}]_{Vir}, \quad \quad \quad [\frac{6}{10}]_{Vir},
\quad \quad \quad [\frac{3}{2}]_{Vir};}
or in $N=1$ terms
\eqn\ns{NS: \quad \quad \quad \quad [0], \quad \quad \quad \quad \quad \quad
[\frac{1}{10}]}
and

\eqn\r{R: \quad \quad \quad \quad [\frac{7}{16}], \quad \quad
\quad \quad \quad \quad [\frac{3}{80}].}
Supersymmetry requires that Ramond vacuum for whole theory
has dimension $\frac{d}{16}=\frac{7}{16}$, and
this leads to the following unitary highest weight representations
of extended chiral algebra in the Ramond ground state (we use the notation
$[\Delta_I,\Delta_r]$ for operators that correspond to
Virasoro highest weights $|\Delta_I,\Delta_r>$ with first
dimension being the dimension of tri-critical
Ising part and the second the
dimension of the remaining Virasoro algebra $T_r$):
\eqn\rr{R: \quad \quad \quad \quad |\frac{7}{16},0>
\quad \quad \quad \quad |\frac{3}{80},\frac{2}{5}>}
It is one of the most remarkable facts for this theory that there
exists a ground state in the Ramond sector which is entirely constructed
out of the tri-critical Ising sector, namely the $|\frac{7}{16},0>$
state.  It is as if the tri-critical Ising model `knows' about the
fact that the dimension of the manifold of interest is 7.  As we will
see this is crucially related to having an $N=1$ spacetime supersymmetry
as well as the possibility of twisting the theory.  In many ways the operator
corresponding to this ground state plays the same role as the spectral
flow operator in $N=2$ theories which is also entirely built out of the
$U(1)$ piece of $N=2$.  To have one spacetime supersymmetry we would be
interested in realization of this algebra which has exactly
 one Ramond ground
state of the form $|{7\over 16},0>$ (we shall make this statement
a little bit more precise when we talk about putting left- and right-movers
together).  In this regard it is crucial to note that in the
tri-critical Ising model
we have unique fusion rules for the operator
$[\frac{7}{16}]$
\eqn\isop{[\frac{7}{16}][\frac{7}{16}]=[0]_{Vir}+[\frac{3}{2}]_{Vir}=[0],}
\eqn\fus{[\frac{7}{16}][\frac{3}{80}]=[\frac{1}{10}]_{Vir}+[\frac{6}{10}]_{Vir}
=[\frac{1}{10}].}
The existence of this operator in the Ramond sector
 allows us to {\it predict} the existence
of certain states in the NS sector.  This follows
from the fact that it sits entirely in the tri-critical Ising part of the
theory and its OPE with other fields depend only on the tri-critical Ising
content of other state and thus
 by considering the OPE
of the operator corresponding to
 $|\frac{7}{16},0>$ state with the other states in the Ramond sector
we end up with certain special NS states.
{}From \isop\ we conclude that
Ising spin field $[\frac{7}{16}]$
maps Ramond ground state $|\frac{7}{16},0>$
to NS vacuum $|0,0>$ and  vice versa.
More importantly when we consider the OPE of the $|\frac{7}{16},0>$
state with $|\frac{3}{80},\frac{2}{5}>$ we end up with a primary state
in the NS sector of the form
$|\frac{1}{10},\frac{2}{5}>$,
which has total dimension $\frac{1}{2}$ and
is primary. This procedure can
be repeated in opposite direction: tri-critical Ising model
spin field $[\frac{7}{16}]$ maps primary
field of NS sector $[\frac{1}{10},\frac{2}{5}]$
to an R ground state $|\frac{3}{80},\frac{2}{5}>$. This leads to
the prediction of existence of the following
special states in NS sector:
\eqn\pic{NS: \quad \quad \quad \quad |0,0>, \quad \quad \quad \quad
|\frac{1}{10},\frac{2}{5}>.}
Note in particular that since the $T_r$ part of the theory is un-modified
as we go from the R sector to the NS sector.  It is again quite remarkable
that the state in the NS sector corresponding to $|\frac{1}{10},\frac{2}{5}>$
is a primary field of dimension 1/2 and so $G_{-1/2}$ acting on it
is of dimension 1, preserving N=1 supersymmetry and thus a candidate
for exactly marginal perturbation in the theory.  We will use
the extended chiral algebra below to show that indeed they lead
to exactly marginal directions.  Again the fact that this state has
dimension 1/2 is a consequence of a miraculous relation between
the dimension of tri-critical Ising model states. If one traces
back one finds that it comes from the fact that
$${7\over 16}-{3\over 80}+{1\over 10}={1\over 2}.$$
In the above discussion we assumed that $Z_2$ fermion
number assignment on any state is equal to the
$Z_2$ grading for its tri-critical part alone which in particular
implies that in the NS sector of the full theory only NS dimensions of
tri-critical model show up and similarly in the R sector.
Let us now discuss how this comes about.
Our chiral algebra has three
bosonic $T, X, K$ and three fermionic $G, \Phi, M$ generators.
 We have the
following tri-critical $Z_2$
 assignments: $[0]^+, [\frac{1}{10}]^-,
[\frac{6}{10}]^+, [\frac{3}{2}]^-$. To prove that $(-1)^F=(-1)^{F_I}$
it suffices to derive tri-critical Ising dimensions of our generators
and see if the two $Z_2$ assignments agree.
Here we have to use relations presented in Appendix 2; we have

\eqn\gener{L_{-2}|0,0>=|2,0>^++|0,2>^+, \quad  X_{-2}|0,0>=|2,0>^+,
\quad  K_{-2}|0,0>=|\frac{6}{10},\frac{14}{10}>^+,}
\eqn\generfer{
G_{-3/2}|0,0>=|\frac{1}{10},\frac{14}{10}>^-,  M_{-5/2}|0,0>=
a|\frac{1}{10},\frac{24}{10}>^-+b|\frac{1}{10}+1,\frac{14}{10}>^-.}
We see that in the assignment in above expressions $(-1)^F=(-1)^{F_I}$
and thus we can use tri-critical gradings for the whole theory.

Now we are ready to discuss the non-chiral, left-right sector.
This will also lead to a better understanding of the correspondence
with geometry.
We claim that only states in (R,R) ground state are:
\eqn\rr{(R,R): \quad \quad
 |(\frac{7}{16},0)_L;(\frac{7}{16},0)_R;\pm >, \quad \quad
|(\frac{3}{80},\frac{2}{5})_L;(\frac{3}{80},\frac{2}{5})_R;\pm >.}
where the significance of $\pm$ will be explained momentarily.
We had two other possibilities of left-right combinations:
$|(\frac{7}{16},0)_L;
(\frac{3}{80},\frac{2}{5})_R;\pm >$ and the same with exchange of
$L$ with $R$. The reason we didn't put these states in the list \rr\ is simple.
If we use fusion rules \isop\ and \fus\ we see that primary field
corresponding to first ground state in \rr\ acting on these additional states
will lead (according to tri-critical
Ising model fusion rules) to the highest weight state
$|(0,0)_L;(\frac{1}{10},
\frac{2}{5})_R>$ in the Neveu-Schwarz sector. But, this operator
has total dimension $\frac{1}{2}$ and is chiral, so, we get an additional
chiral
operator of half-integer spin in the theory which is not present in our
original extended chiral algebra. This means that these additional
states aren't present in the case of generic theory (which is assumed to have
only chiral operators described in the beginning of this section).
 The $\pm$ signs next to the states are a reflection of the
fact that
since acting on the ground states we  have
$\{\Phi_0,{\bar \Phi_0}\}=0, {\Phi}_0^2={\bar \Phi}_0^2=\frac{6}{15}$,
they form a 2 dimensional representation.  The $\pm$ sign therefore
reflects states with 2 different $(-1)^F$ assignments.
Thus, Ramond ground states are
coming
in pairs - $\Phi_0|(\frac{7}{16},0,)_L;(\frac{7}{16},0)_R;+>=
|(\frac{7}{16},0)_L;(\frac{7}{16},0)_R;->,
\Phi_0|(\frac{3}{80},\frac{2}{5})_L;(\frac{3}{80},\frac{2}{5})_R;+>=
|(\frac{3}{80},\frac{2}{5})_L;(\frac{3}{80},\frac{2}{5})_R;-)>$.
Now we can better describe the relation of Ramond ground states with
the cohomology of the manifold:  The fact that states come in pairs
is a consequence of the fact that in odd dimension the dual of every
cohomology state is another cohomology state with different  degree
mod 2. So the Ramond $+$ states correspond to even cohomology elements
and $-$ to the odd ones.
 So now concentrating on the even cohomology elements in principle
we could have $b_0=1,b_2, b_4$ as the elements (note that having
no extra supersymmetry leads to having $b_6=b_1=0$ which is
correlated with the fact that we assume the $|(\frac{7}{16},0)_L;
(\frac{7}{16},0)_R,+>$ is unique).  We see that we can only compute
one extra number, and not two, which is the number of ground states
involving the ${3\over 80}$ tri-critical piece for both left- and right-movers
which we identify with $b_2+b_4$.

Let us discuss the special NS states taking into account both the left- and
right-moving degrees of freedom. Acting on all
$+$ Ramond ground states with the state $|(\frac{7}{16},0)_L;
(\frac{7}{16},0)_R,+>$
 leads to $(NS,NS)$ states
\eqn\nsrmap{(NS,NS): \quad \quad \quad \quad |(0,0)_L;(0,0)_R>
\quad \quad
\quad \quad \quad \quad
|(\frac{1}{10},\frac{2}{5})_L;(\frac{1}{10},\frac{2}{5})_R>.}
where the number of
$|(\frac{1}{10},\frac{2}{5})_L;(\frac{1}{10},\frac{2}{5})_R>$
states are the same as the states
$|(\frac{3}{80},\frac{2}{5})_L;(\frac{3}{80},\frac{2}{5})_R>$
which is equal to $b_2+b_4$.\foot{Also in principle we will get
other higher dimension states such as
$|{3\over 2},0)_L;({3\over 2},0)_R>$ or $|({6\over 10},{2\over 5})_L;
({6\over 10},{2\over 5})_R>$.}
  Moreover as we will argue
later in this section each of
all such NS operators are exactly marginal operators
preserving the $G_2$ structure.  This agrees with the geometrical
facts discussed in section 2 in that the dimension of conformal
moduli space is thus $b_2+b_4=b_2+b_3$.

Before we address the question of marginal deformations of our conformal field
theory let us discuss the relation of the above construction to
10-dimensional Superstring Theory compactified down to 3-dimensions.
It is easy to show that if corresponding compact 7-dimensional manifold
is a $G_2$-manifold we will have $N=2$ supersymmetry
for type II strings and $N=1$ supersymmetry for heterotic strings
 in 3-dimension.
Let us construct the corresponding supersymmetry generators using all
the information that we already obtained. We have:

\eqn\super{J_{L,R}=e^{-\frac{\phi_{gh}}{2}}S_3^{\alpha}
\sigma_{\frac{7}{16}}^{L,R}.}
Here $\phi_{gh}$ is a bosonized 10-dimensional ghost field, $S^{\alpha}_3$
are 3-dimensional spin fields and $\sigma$ is tri-critical Ising model
spin field that we had already discussed many times.\foot{This is
a standard ansatz for target space supersymmetry current, see \ref\emil{
T. Banks, L. Dixon, D. Friedan and E. Martinec, Nucl. Phys., B299 (1988)
613.}.}
First we notice that $J$ has dimension 1; dimension of 10-dimensional
ghost part doesn't depends on compactification and always is equal to
$\frac{3}{8}$, dimension of 3d spin field is $3.\frac{1}{16}=\frac{3}{16}$ and
dimension of sigma by definition is $\frac{7}{16}$, and all add up to 1.
If we remember that $\sigma$ has a unique OPE with vacuum $[0]$ in
the right hand side we can consider $J$ as a chiral operator and this explains
subscript $L,R$ in \super. Now we can define 3d supersymmetry generators:
$Q_{L,R}={\oint}J_{L,R}$ and standard computation leads to
supersymmetry algebra.
Also, one finds that
one of the supersymmetry transforms of
$(\frac{3}{80},\frac{2}{5},)_{L,R}$ which is accompanied by
spacetime spinor field and ghost degrees of freedom is simply the state
$(\frac{1}{10},\frac{2}{5})_{L,R}$.

Now we would like to consider marginal deformations of our
theory.  As mentioned before we will show
that marginal deformations are given by perturbation with dimension 1
operators of the form
$G^L_{-1/2}G_{-1/2}^R[(1/10,2/5)_L;(1/10,2/5)_R]$;
the dimension of this moduli space is $b^2+b^3$. In addition
to showing that they preserve $N=1$ superconformal symmetry
we need to show that they do not have any tri-critical
piece in them, which would otherwise destroy the
existence of the extended algebra in question.  This
follows because the full algebra was generated by the
$N=1$ algebra together with the supersymmetry operator
$\Phi$ of the tri-critical model.  We will first show
this fact by studying the
content of above operator with respect to tri-critical Ising model.
For this we have to apply the operator $X_0$.
We have (it is enough to consider only chiral sector):

\eqn\ising{\eqalign{X_0G_{-1/2}&|\frac{1}{10},\frac{2}{5}>=
G_{-1/2}X_0|\frac{1}{10},\frac{2}{5}>+
[X_0,G_{-1/2}]|\frac{1}{10},\frac{2}{5}>=\cr
&(-\frac{1}{2}G_{-1/2}-M_{-1/2})|\frac{1}{10},\frac{2}{5}>=P.}}
It turns out that the right hand side of this equation
is identically zero in our
theory: $P=0$. One can check that this state
has zero norm and $P$ is a null vector:

\eqn\norm{\eqalign{|P|^2=&<\frac{1}{10},\frac{2}{5}|
(G_{1/2}-M_{1/2})(\frac{1}{2}G_{-1/2}+M_{-1/2})|\frac{1}{10},\frac{2}{5}>
=\cr
&<\frac{1}{10},\frac{2}{5}|(2L_0-2X_0+8X_0L_0)|\frac{1}{10},\frac{2}{5}>
=0.}}
Here we used the fact that $|\frac{1}{10},\frac{2}{5}>$
is a highest weight representation of the whole extended chiral algebra,
the property $M_n^+=\frac{1}{2}G_{-n}-M_{-n}$
the commutation relations given in Appendix 1 and
$2L_0|\frac{1}{10},\frac{2}{5}>=
-2X_0|\frac{1}{10},\frac{2}{5}>=|\frac{1}{10},\frac{2}{5}>.$
So, we conclude that our deformation is of the type $[(0,1)_L;(0,1)_R]$.
So all we are left to show is that the deformation preserves
conformal invariance.

For simplicity we will denote our perturbation
by $G^L_{-1/2}A(z,\bar z)$ (we will work with the chiral part
below and thus will suppress $\bar z$ dependence and $G^R_{-1/2}$).
The following proof is based on two facts:

1.  Dixon \dix\ has shown using just N=1 superconformal algebra that
perturbation with dimension 1 operator of the form $G_{-1/2}A$
is marginal if

\eqn\marg{F = <G_{-1/2}A(z_1)G_{-1/2}A(z_2)G_{-1/2}A(z_3)G_{-1/2}A(z_4)...
G_{-1/2}A(z_n)>}
is a total derivative with respect to coordinates $z_i$,  $i>3$.
Perturbation is truly marginal if n-point
correlation function \marg\ integrated over all points, except
the first three, is
zero (first three points are fixed by $SL(2,C)$ invariance on sphere) and
Dixon has shown that in N=1 super conformal theory
the integrand can be regulated in such a fashion that if it
is a total derivative there are no contact term contributions.\foot{If there is
a total derivative in holomorphic variable by symmetry we get
total derivative both in holomorphic and antiholomorphic
coordinates ${\partial}_{z_{i}}{\partial}_{{\bar z}_j}$
and this
is crucial in showing that there are no contact term
contributions.}

2. As we have seen above $A(z)$ has a null vector

\eqn\nu{(\frac{1}{2}G_{-1/2} + M_{-1/2})A(z) = 0;}
In addition we need several relations between the generators of
the extended
algebra acting on $A(z)$ (which is a highest weight vector and thus
is killed by positive energy modes of all generators):

\eqn\rel{M_{1/2}G_{-1/2}A(z) = -2X_0A(z) = A(z),}
\eqn\rele{M_{-1/2}G_{-1/2}A(z) = (- X_{-1} + \frac{1}{2}L_{-1})A(z),}
\eqn\relet{M_{-3/2}G_{-1/2}A(z) = - L_{-1}X_{-1}A(z).}

In fact, one can show that it is enough to prove that

\eqn\i{I_0 = <G_{-1/2}A(z_1)A(z_2)A(z_3)G_{-1/2}A(z_4)...G_{-1/2}A(z_n)>}
is a total derivative $\frac{\partial}{{\partial}z_i}$, $i>3$,
of something. For this we need to write $G_{-1/2}A(z_1)
= {\oint}_{z_{1}}G(z)dzA(z_1)$ in \marg\ and deform the contour.
If we remember that the vacuum
is annihilated by $G_{+1/2}$ and  $G_{-1/2}$ (and also by
$M_k, k = 3/2, 1/2, -1/2, -3/2$; this we will need later) we will
have no contribution from infinity and:

\eqn\garb{\eqalign{F =
& - <A(z_1)L_{-1}A(z_2)G_{-1/2}A(z_3)G_{-1/2}A(z_4)...G_{-1/2}A_(z_n)>
-\cr
&<A(z_1)G_{-1/2}A(z_2)L_{-1}A(z_3)G_{-1/2}A(z_4)...G_{-1/2}A(z_n)> - ... =\cr
&- \frac{\partial}{{\partial}z_2}<A_(z_1)A(z_2)G_{-1/2}A(z_3)G_{-1/2}A(z_4)
...G_{-1/2}A(z_n)> -\cr
&\frac{\partial}{{\partial}z_3}<A(z_1)G_{-1/2}A(z_2)A(z_3)
G_{-1/2}A(z_4)...G_{-1/2}A(z_n)> - ...,}}
where we dropped the terms that are total derivatives
with respect to $z_i, i >3$: $G_{-1/2}^2=L_{-1}={\partial}_z$.
Thus, if we can show that $I_0$ is zero modulo $\frac{\partial}{{\partial}z_i},
i>3$ of something, we will have the proof of marginality: ${\int}F = 0$.
Below we will deal with the object $I = {{\int}}d^2z_4...d^2z_nI_0$ and
ignore total derivatives inside the integral referring the reader to the
regularization used by Dixon.

Our main strategy is to use the null vector condition \nu\ and contour
deformation argument first for $G_{-1/2}A(z_1)$ in $I$ and then
the same argument but now replacing
$G_{-1/2}A(z_1)$ by $-2M_{-1/2}A(z_1)$. First we insert
${\oint}_{\infty}(w-z_l)G(w)$ with contour around infinity in the
correlator $<A(z_1)A(z_2)A(z_3)({\int}G_{-1/2}A(z))^{n-3}>$ and
place the zero $z_l$ at $z_3$ and $z_2$. After the contour deformation we get:

\eqn\iden{\eqalign{(z_1-z_3)&
<G_{-1/2}A(z_1)A(z_2)A(z_3)({\int}G_{-1/2}A(z))^{n-3}> +\cr
&(z_2-z_3)<A(z_1)G_{-1/2}A(z_2)A(z_3)({\int}G_{-1/2}A(z))^{n-3}> = 0,}}
\eqn\ident{\eqalign{(z_1-z_2)&<G_{-1/2}A(z_1)A(z_2)A(z_3)
({\int}G_{-1/2}A(z))^{n-3})> +\cr
&(z_3-z_2)<A(z_1)A(z_2)G_{-1/2}A(z_3)({\int}G_{-1/2}A(z))^{n-3}> = 0.}}
Here we used the mode expansion

\eqn\mode{{\oint}_{z_k}(z-z_l)G(z)B(z_k)= ((z_k-z_l)G_{-1/2} + G_{1/2})B(z_k)}
for any $B$.  The
total derivative term that was ignored has an insertion of
\eqn\inser{{\int}d^2z_4[(z_4-z_l)L_{-1}+2L_0]A(z_4)
= {\int}d^2z_4\frac{\partial}{{\partial}_{z_4}}(z_4-z_l)A(z_4)}
and this identity holds only if $A$ has dimension $\frac{1}{2}$: $2L_0A=A$.

A similar formula can be written for $M$, which has dimension 5/2, and thus
we need to insert ${\oint}_{\infty}v(z)M(z)$ with $v$
now having three zeros. Placing
zeros at points $z_1, z_2, z_3$ we get:

\eqn\mo{\eqalign{{\oint}_{z_k}&(z-z_1)(z-z_2)(z -z_3)M(z)B(z_k) =
[M_{3/2} + (z_k - z_1 + z_k - z_2 + z_k -z_3)M_{1/2} +\cr
&((z_k-z_1)(z_k-z_2) +
(z_k-z_1)(z_k-z_3) + (z_k-z_2)(z_k-z_3)]M_{-1/2} +\cr
&(z_k-z_1)(z_k-z_2)(z_k-z_3)M_{-3/2}]B(z_k).}}

Now we consider correlation function:

\eqn\main{<{\oint}_{\infty}(w-z_1)(w-z_2)(w-z_3)M(w)A(z_1)A(z_2)A(z_3)
({\int}G_{-1/2}A(z))^{n-3}>=0,}
with contour around infinity. Again, because all modes of $M$ that
enter in \mo\ kill the vacuum, the right hand side of \main\ is zero;
we could deform the contour and obtain the identity:

\eqn\mainn{\eqalign{&(z_1-z_2)(z_1-z_3)<M_{-1/2}
A(z_1)A(z_2)A(z_3)({\int}G_{-1/2}A(z))^{n-3}> +\cr
&(z_2-z_1)(z_2-z_3)<A(z_1)M_{-1/2}
A(z_2)A(z_3)({\int}G_{-1/2}A(z))^{n-3}> +\cr
&(z_3-z_1)(z_3-z_2)<A(z_1)A(z_2)M_{-1/2}A(z_3)
({\int}G_{-1/2}A(z))^{n-3}> +\cr
&(n-3)<A(z_1)A(z_2)A(z_3){{\int}}
d^2z_4(z_4-z_1+z_4-z_2+z_4-z_3)M_{1/2}G_{-1/2}A(z_4)\cr
&({\int}G_{-1/2}A(z))^{n-4}> +
(n-3)<A(z_1)A(z_2)A(z_3){{\int}}d^2z_4[(z_4-z_1)(z_4-z_2) +\cr
&(z_4-z_1)(z_4-z_3)
+ (z_4-z_2)(z_4-z_3)]M_{-1/2}G_{-1/2}A(z_4)({\int}G_{-1/2}A(z))^{n-4}>
+\cr
&(n-3)<A(z_1)A(z_2)A(z_3){{\int}}d^2z_4
(z_4-z_1)(z_4-z_2)(z_4-z_3)\cr
&M_{-3/2}G_{-1/2}A(z_4)({\int}G_{-1/2}A(z))^{n-4}> = 0.}}
Now we use relations \rel, \rele\ and \relet, and simply find that the
last three terms combined lead to the integral
of total derivative in $z_4$. More concretely, we write $L_{-1}={\partial}$
and integrating by part in last term of \mainn\ using \relet\
we cancel contribution
of $X_{-1}$ from \rele\ in
the previous term; similarly, after integration by parts, second
term from \rele\ kills the contribution of $X_0$ from \rel.
\foot{
The terms that have been ignored here are
total derivatives only if $2L_0A=-2X_0A=A$,
and this condition is exactly satisfied by our choice of $A$.}
 Thus, we drop
these terms and replace $M_{-1/2}$ by $-\frac{1}{2}G_{-1/2}$.
Combined with the
identities \iden\ and \ident\ we see that ${-\frac{3}{2}}I = 0$.
This leads to
the proof of the statement that our perturbation is truly marginal.
It is very satisfying that we used many different aspects of the
extended chiral algebra for this proof.

\subsec{$Spin(7)$}

We will follow the ideas described above for the case of $G_2$-holonomy
and first discuss extended symmetry for the sigma models on $Spin(7)$
manifolds. This story is completely parallel to the previous case
and so we will be brief.

First we describe symmetry algebra in free field representation. As before,
we take $Spin(7)$ 4-form and replace $e$ by target space fermions; thus
we get a spin 2 operator - $\tilde X$:

\eqn\pphi{\tilde X = \psi^8\Phi - X + 1/2 {\partial}\psi^8\psi^8.}
Pleasantly we find that the
 operator $T_I=\frac{1}{8}\tilde X$ forms a
Virasoro algebra
with central charge $\frac{1}{2}$ and this means that the tri-critical Ising
model that we had in the
previous case is replaced by the ordinary, bosonic Ising
model as predicted at the beginning of this section.
As before, we have to check
operator expansion with original $N=1$ generators and we immediately
find that $\tilde X$ has a super partner - $\tilde M$:
\eqn\Gpphi{G(z)\tilde X(w) = 1/2(z-w)^2G(w) + 1/(z-w)\tilde M(w),}
with
\eqn\tt{\tilde M = J^8\Phi - \psi^8K - M +1/2{\partial}J^8\psi^8 -
1/2 J^8{\partial}\psi^8.}
This operator has dimension $\frac{5}{2}$ and will play the role of the
operator $M$. It turns out that these four operators,
$G, T, \tilde X$ and $\tilde M$, form a closed operator expansion algebra,
which again is a quadratic $W$-type algebra. Corresponding formulas together
with mode expansion are given in the Appendix 1. From this extended symmetry
algebra it follows that one can again
decompose original stress-tensor as a
sum of two commutative Virasoro generators:
\eqn\decom{T=T_I+T_r,}
and we can classify our states again by two numbers: Ising model highest weight
and the eigenvalue of the zero mode of $T_r$: $|\Delta_I,\Delta_r>$.

In chiral (left-mover) sector above observation immediately leads to
the following content:

\eqn\spinns{ |0,\Delta_r> \quad \quad \quad \quad
|\frac{1}{2},\Delta_r> \quad \quad \quad \quad |\frac{1}{16},\Delta_r>.}
This means that in the Ramond sector, where we have to have dimension of
ground state equal to $\frac{8}{16}=\frac{1}{2}$,
(this follows from the requirement
of supersymmetry - dimension of the Ramond ground state has to be equal to
$\frac{c}{24}$) we should have the following highest
weight states:
\eqn\ramonspin{R: \quad \quad \quad \quad |\frac{1}{2},0>
\quad \quad \quad \quad |0,\frac{1}{2}> \quad \quad \quad \quad
|\frac{1}{16},\frac{7}{16}> .}
Amazingly enough there is again a unique state in the ground
state built purely from the Ising piece, which is the
$|{1\over 2},0>$ state.  This will now play an identical role to that of
spin operator of tri-critical Ising model
$[\frac{7}{16}]$ that mapped Ramond ground state to NS sector and
vice versa; the specific property this operator had was that it had unique
fusion rules
with itself and other operator from Ramond ground state. In the $Spin(7)$
model this operator is replaced by the Ising model energy operator
$\epsilon = [\frac{1}{2}]$; it has unique fusion rules and maps the Ramond
ground state to a certain special NS highest weight states and vice versa:

\eqn\nsspin{NS: \quad \quad \quad \quad |0,0> \quad \quad \quad \quad
|\frac{1}{2},\frac{1}{2}> \quad \quad \quad \quad |\frac{1}{16},\frac{7}{16}>
.}
Here we are using Ising model fusion rules: $[\epsilon][\epsilon]=[0],
[\epsilon][\sigma]=[\sigma], [\sigma][\sigma]=[0]+[\epsilon],
[\sigma]=[\frac{1}{16}].$ The operator $(\frac{1}{16},\frac{7}{16})$ has
total dimension $\frac{1}{2}$ and clearly is a candidate for marginal
deformation after acting by $G_{-1/2}$ on it. Again the fact that the
dimension of this operator is ${1\over 2}$ is magical
and related to the existence of spacetime supersymmetry.

In the Ising sector we have $Z_2$ symmetry: $\sigma \rightarrow -\sigma;
1,\epsilon \rightarrow 1,\epsilon$. We would like to show that corresponding
$(-1)^{F_I}$ is again identified with total $(-1)^F$.  As in
the $G_2$ case we
have
to compute Ising content of the generators of the chiral algebra. We have:

\eqn\gengen{L_{-2}|0,0>=|2,0>^++|(0,2>^+, \tilde X_{-2}|0,0>=|2,0>^+,}
\eqn\genfermion{G_{-3/2}|0,0>=|\frac{1}{16},\frac{23}{16}>^-,
\tilde M_{-5/2}|0,0>=a|\frac{1}{16}+1,\frac{23}{16}>^-+
b|\frac{1}{16},\frac{39}{16}>^-,}
and we had used the commutation relations from Appendix 1. Now we see that
$(-1)^{F_I}=(-1)^F$. Thus we use Ising model fermion number assignment.

Let us now discuss non-chiral sector
putting left and right sectors together. We claim that the content of
RR ground state is given by the following combinations:

\eqn\rrspin{\eqalign{RR:
&\quad \quad \quad |(\frac{1}{2},0)_L;(\frac{1}{2},0)_R>,
\quad \quad \quad \quad \quad \quad
|(0,\frac{1}{2})_L;(0,\frac{1}{2})_R>,\cr
& \quad |(0,\frac{1}{2})_L;(\frac{1}{16},
\frac{7}{16})_R>, \quad |(\frac{1}{16},\frac{7}{16})_L;(0,\frac{1}{2})_R>
\quad |(\frac{1}{16},\frac{7}{16})_L;(\frac{1}{16},\frac{7}{16})_R>.}}
Other possible combinations can be ruled out by similar arguments as in
the
$G_2$ case -- using Ising model fusion rules they lead to existence of chiral
half-integer spin operators that are not present in extended chiral algebra
and thus such combinations can't appear in the ground state of a
generic model.

We now wish to connect the above states as much as possible with the
cohomology of the manifold.  As far as even degrees are concerned
they come from first, second and last state which all have $(-1)^F=+1$.
  Moreover we will connect all the
NS versions of the last state with
exactly marginal deformations, and so as discussed
in section 2 there are $1+b_2+b_4^-$ of them.
Moreover the condition of having exactly one supersymmetry means that
the first state is unique.
So the second states are as many as
$b_6+b_4^+$.
The second and third state correspond to odd cohomology
elements and each one are in number equal to $b_3=b_5$.

Using the unique analog of spectral flow the
above content of (R,R) ground state after mapping to (NS,NS) sector due to
Ising model energy operator leads to following special states

\eqn\nsnsspin{\eqalign{(NS,NS): &\quad \quad \quad |(0,0)_L;(0,0)_R>,
\quad \quad \quad \quad \quad \quad
|(\frac{1}{2},\frac{1}{2})_L;(\frac{1}{2},\frac{1}{2})_R>,
\quad \cr
&|(\frac{1}{2},\frac{1}{2})_L;(\frac{1}{16},\frac{7}{16})_R>, \quad
|(\frac{1}{16},\frac{7}{16})_L;(\frac{1}{2},\frac{1}{2})_R>, \quad
|(\frac{1}{16},\frac{7}{16})_L;(\frac{1}{16},\frac{7}{16})_R>.}}

As we already mentioned operator $G^L_{-1/2}G^R_{-1/2}
[(\frac{1}{16},\frac{7}{16})_L;
(\frac{1}{16},\frac{7}{16})_R]$ is a candidate for marginal perturbation.
Again we wish to show that the Ising structure is not affected
by this perturbation.  In other words we will show that
this operator has zero dimension in Ising part. To demonstrate this
fact we have to show that it is annihilated by $\tilde X_0$ (again we will
keep only chiral part in this computation):

\eqn\margdim{\eqalign{\tilde X_0G_{-1/2}&|\frac{1}{16},\frac{7}{16}>=
G_{-1/2}\tilde X_0|\frac{1}{16},\frac{7}{16}>+[\tilde X_0,G_{-1/2}]
|\frac{1}{16},\frac{7}{16}>=\cr
&(\frac{1}{2}G_{-1/2}-\tilde M_{-1/2})|\frac{1}{16},\frac{7}{16}>=\tilde P.}}
$\tilde P$ is a null vector, $\tilde P=0$,
similar to the one in $G_2$ case \nu.
We have for norm:

\eqn\normm{\eqalign{|\tilde P|^2=<\frac{1}{16},&\frac{7}{16}|
(G_{1/2}+\tilde M_{1/2})(\frac{1}{2}G_{-1/2}-
\tilde M_{-1/2})|\frac{1}{16},\frac{7}{16}>=\cr
&<\frac{1}{16},\frac{7}{16}|(2L_0-8\tilde X_0+12{\tilde X_0}L_0)
|\frac{1}{16},\frac{7}{16}>=0.}}
We used commutation relations given in Appendix 1 and relations:
$\tilde M^+_n=-\frac{1}{2}G_{-n}-\tilde M_{-n}$,
$2L_0|\frac{1}{16},\frac{7}{16}>=2\tilde X_0|\frac{1}{16},\frac{7}{16}>=
|\frac{1}{16},\frac{7}{16}>$.
So, we see that $G_{-1/2}[\frac{1}{16},\frac{7}{16}]$ is of the type $(0,1)$
and if it is truly marginal
it will preserve also extended $Spin(7)$ symmetry.
In addition we got a very important null
vector that will allow us to
prove exact marginality as in the case of $G_2$.

In fact, the only information
from extended chiral algebra we
had used in
the $G_2$
case to prove exact marginality was null vector condition (relation
between $G_{-1/2}A$ and $M_{-1/2}A$) and commutation relation \rel, \rele,
\relet. Null vector condition $\tilde P=0$ is practically the same
(relative coefficient in $\tilde P$ doesn't play a key role) and analog
of \rel, \rele, \relet\ can be derived from the expressions in the Appendix 1:
\eqn\rell{{\tilde M}_{1/2}G_{-1/2}A=-2{\tilde X_0}A=-A,}
\eqn\relel{{\tilde M}_{-1/2}G_{-1/2}A=(-\frac{1}{2}L_{-1}-{\tilde X_{-1}})A,}
\eqn\relete{{\tilde M}_{-3/2}G_{-1/2}A=-L_{-1}{\tilde X_{-1}}A;}
we use the notation $A=G_{-1/2}^R[(\frac{1}{16},\frac{7}{16})_L;
(\frac{1}{16},\frac{7}{16})_R]$.
Now the argument
presented in the case of $G_2$ can be repeated
identically with the same
conclusion-- our perturbation is truly marginal to all orders.

\newsec{Examples of Joyce}

Here we will study some of the examples constructed by Joyce
\joyce .   We will review his description
of some of his models.
It will be clear from the construction that the
story is easily generalizable using the standard methods
familiar from orbifold constructions \ref\orbi{L. Dixon, J. Harvey, C. Vafa
and E. Witten, Nucl. Phys. B261 (1985) 678, Nucl. Phys. B274 (1986) 285.}.
Let us discuss a $G_2$ example first (example 4 of II in \joyce ):
Consider $T^7$ modded out by $Z_2^3$ where the generators
of the $Z_2$'s we denote by $\alpha , \beta , \gamma $.
Let us represent each of them
by a pair of row vectors: the holonomy part of these
elements, which are simultaneously diagonal, by a row
of 7 ($\pm 1$)'s  and they are accompanied by shifts
acting as translation on the torus which
again is written by another row vector.  We take each
of the 7 coordinates $x_i$ of $T^7$ to have period 1.
Then
$$\alpha =[(-1,-1,-1,-1,1,1,1);(0,0,0,0,0,0,0)],$$
$$\beta =[(-1,-1,1,1,-1,-1,1);(0,{1\over 2},0,0,0,0,0)],$$
$$\gamma =[(-1,1,-1,1,-1,1,-1);({1\over 2},0,0,0,0,0,0)].$$
Note that the above holonomies preserve $\phi$ defined in \ffi ,
and that they do not sit in an $SU(3)$ group as there is no
invariant direction.  If we look at the untwisted Ramond sector,
which can be identified with the cohomology elements of the torus,
we see that of the cohomologies of the torus we project out all
except for the $H^0$,$H^7$ which
are one dimensional and $7$ in $H^3$ and $7$ in $H^4$.  The 7 invariant
elements precisely correspond to the 7 monomials in the definition
of the forms $\phi$ \ffi\ and $*\phi $ \dfi .  It is straightforward
to construct the 7 twisted sectors. However
since we are interested in the topological aspects, let us concentrate
on the sectors which give rise to new ground states in the
Ramond sector.  For this to happen
there should be fixed points for the group action.  It is easily
seen that out of the 7 non-trivial elements only three have
fixed points, namely $\alpha ,\beta$ and $\gamma$.  The fixed point
set of $\alpha$ consists of $2^4$ three tori,
each of which has 8 cohomology elements $(1,3,3,1)$.  To get the final answer
we have to
project to the invariant subsector under the action of the full group.
$\beta$ and $\gamma$ act freely on this set and
leave us with $4$ invariant combinations of the
16 $T^3$'s.  So finally we have 4 copies of $(1,3,3,1)$
added to the Ramond ground state from this sector. Similarly
one can easily see that from the $\beta$ sector
after projection we get 4 copies of $T^3$.  As far
as the structure discussed in the previous section is concerned
we can only say that we get a contribution to the $b_2+b_4=4$
and to the $b_3+b_5= 4$
from each of the total of 8 tori coming from the $\alpha$ and $\beta$
sector.
The $\gamma$
sector projected to its invariant fixed point set gives
8 copies of $T^3/Z_2$, where the $Z_2$ acts in the
neighborhood of each of these $T^3$'s
by
$$(y_1,y_2,y_3,z_1,z_2)\rightarrow ({1\over 2}+y_1,-y_2,-y_3,z_1,-z_2),$$
where $y_i$ denote the coordinates of the fixed $T^3$ and the $z_i$
denote in complex notation the orthogonal direction (which by the
action of $\gamma$ goes to minus itself).  Of the cohomologies
of each of these $T^3$'s from the above $Z_2$ action only two
elements survive, two in odd and two in even cohomology,
so we get from the total of 8 tori the addition of 16 to $b_2+b_4$
and addition of 16 to $b_3+b_5$ from
the $\gamma$ sector.  If we put the contributions
of all the sectors together we find
$$b_0=1 \semi b_2+b_4=55 \semi b_3+b_5=55 \semi b_7=1.$$

As noted in the previous section we must thus have a 55 dimensional
moduli space:  7 of the moduli come from the untwisted sector
and correspond to the 7 radii of $T^7$. The other 48 come from
blow up modes in the twisted sectors.  As proven in the previous
section all these deformations are exactly marginal.  Just to give
 a better feeling for how the algebra discussed in the previous
section fit with the geometry let us describe the untwisted moduli.
The primary superconformal field of dimension 1/2 which
correspond to the untwisted moduli are nothing but the $\psi^i$
for $i=1,..,7$.  From equation \c\  we see that $\psi^i$ has
under $X_0$ the eigenvalue $-1/2$ which implies that for the
tri-critical part of the energy momentum tensor it has eigenvalue
$-X_0/5=1/10$ as predicted by the analysis in the previous section.
Note that we see the crucial role played by the normal ordered terms
in the definition of $X$, which is responsible for giving $\psi^i$
a tri-critical dimension of $1/10$ rather than zero.  Also note that
when we take $G_{-1/2}\psi^i=\partial X^i =J^i$ and it is easy
to see that it thus commutes with $X_0$.  This in particular means
that the tri-critical dimension of it is $0$, again a fact proven
in full generality in the previous section.

Note that as emphasized in the previous section physically
we cannot identify $b_2$ and $b_4$ separately.
Amazingly
enough this structure is reflected mathematically and gives
a first non-trivial example of our generalized mirror conjecture:
there are inequivalent ways the singularities can be resolved
to give manifolds with different betti numbers, but in all these
cases $b_2+b_4$ is the same. More precisely Joyce found that
depending on how he desingularizes the manifold
$$b_2=8+l \semi b_4=47-l,$$
where $l$ runs from $0$ to $8$.  These different
ways of resolving the singularity have to do with the fact
that when one desingularizes the fixed tori of $\gamma$ action
there are different ways that the $Z_2$ that we have to mod out acts:
more precisely, the desingularization  can take place using
the Eguchi-Hanson space which is $T^*(CP^1)$ (as the orthogonal
direction is locally $R^4/Z_2$).  But the $Z_2$ written above
can act in two different ways on the resolved space.  If we let $z$
 be the coordinate of $CP^1$, then the involution acting
on $T^*(CP^1)$ can come from $z\rightarrow - z$
or $z\rightarrow \overline z$. In the first case we get a contribution
to $\Delta b_2=1$ and $\Delta b_3=1$ and in the other case we get
$\Delta b_2=0$ and $\Delta b_3=2$.
  In either case in the limit of shrinking
down the sphere we get the $Z_2$ action above after appropriate
redefinition of coordinates.  It turns out that even though
there are 2 ways of doing the desingularization for each
of the eight tori there are only 9 inequivalent
betti numbers one gets which are listed above.  But we
know physically (from the conformal theory perturbations
discussed in the previous section)
that the moduli space is smooth near the orbifold
point and so at most the difference between these answers
have to do with turning on different marginal operators.  Thus we
see that topologically distinct manifolds, as allowed from the
ambiguity of decoding $b_2$ and $b_4$ give rise to the same
conformal theory (up to moduli deformation) as suggested
by the generalized mirror conjecture\foot{One may be tempted
to identify this with the flop phenomenon for which
distinct manifolds (albeit with the same hodge numbers) are
part of the same moduli space of conformal theory
\ref\green{P. Aspinwall, B. Greene and D. Morrison, Phys. Lett., B303 (1993)
249; Nucl. Phys.,
B416 (1994) 414}\ref\wit{E. Witten, Nucl. Phys., B403 (1993) 159.}.
However, even though there are analogs of flop phenomenon
for $G_2$ manifolds, this is not one of them \ref\joy{D. Joyce,
Private communication.}.}.

Actually there is one subtlety which needs to be considered:
we have assumed that there is a unique orbifold theory.
However there is the possibility of turning on discrete torsion \ref\v{
C. Vafa, Nucl. Phys. B273 (1986) 592.}\ and thus we could have
inequivalent orbifold theories.  In the above
example we could for example turn on a discrete torsion between two of the
$Z_2$'s. However in the case of $Z_2$ torsions this does not
lead to a new theory (and in the case of Calabi-Yau gives
a simple example of mirror symmetry).
  However if instead of $Z_2$'s we had
$Z_n$'s the story would have been different.  Indeed in that case we
expect inequivalent theories at the orbifold points related
to each other by turning on a discrete torsion.  In such
a case one would also expect that geometrically there
should exist inequivalent ways of resolving the singularity--but
here one would not expect them to preserve $b_2 +b_4$ because the
underlying conformal theories are different.  This prediction
has been confirmed by a local model for the $Z_n \times Z_n$
singularity replacing the $Z_2 \times Z_2$ above \joy .
In that
case he finds that there are $n-1$ different choices
of resolution which
lead to $\Delta b_2 =0, \Delta b_4= 2$ and one choice where
$\Delta b_2=\Delta b_4=n-1$.  It is easy to check in the conformal
theory computation that the turning on of the $n-1$ different
possibilities for discrete torsion lead to the first answer and
no discrete torsion leads to the second answer, thus confirming
the correspondence between conformal theory and the geometry
of $G_2$ holonomy manifold.

There are other classes of examples
of $G_2$ holonomy manifolds constructed by Joyce.  One particularly
general construction he suggests is to start
from a Calabi-Yau three fold $M$ which has a real involution
(an involution which locally looks like $z\rightarrow z^*$).
This would be the case for example if one considers
algebraic varieties with real coefficients in the defining
equations.  Then one may obtain a $G_2$ holonomy orbifold by considering
$${M\times S^1\over Z_2},$$
where $Z_2$ sends $M\rightarrow M^*$ and is a reflection on the circle.
It is clear that the holonomy of this $Z_2$ (4 (-1)'s and 2(+1)'s) preserves
supersymmetry and thus lead to a $G_2$ holonomy manifold.
There are orbifold singularities which as we know physically
are harmless.  It is tempting to speculate that using
this construction one can interpolate between Calabi-Yau mirrors
by going through points on the moduli space of the $G_2$ holonomy
manifold where $b_2$ and $b_3$ change but their sum does not
change.

The examples of Spin(7) holonomy manifolds proceeds very
similarly to the above, and so we just summarize the main features.
Again one starts with an $8$ dimensional torus and mods out by
some isometries, the simplest of which is $Z_2^4$
 and resolves the
singularities to obtain a smooth
8 dimensional manifold of $Spin(7)$ holonomy.  Again one sees that there
are inequivalent ways to desingularize manifolds but all have
the property that they lead to the same sum for the even cohomology
elements and for the odd cohomology elements, as predicted from
the conformal theory view point.  These examples therefore provide
further evidence for the generalized mirror conjecture.

\newsec{Topological Twist}

In previous sections we have shown that $G_2$ and $Spin(7)$ compactifications
are very similar to $N=2$ superconformal theories corresponding to
$SU(n)$ or $N=4$ corresponding to $Sp(n)$ holonomy.
   In particular they
both lead to $N=1$ spacetime supersymmetry upon heterotic compactification.
 In $N=2$ (and similarly in the $N=4$  \ref\berv{N. Berkovits
and C. Vafa, in preparation.}) there is a topological side to the story,
which is deeply connected to spacetime supersymmetry in the compactified
theory.  Basically the spectral flow operator, which is the same
operator used to construct spacetime supersymmetry operator is responsible
for the twisting. Twisting is basically the same as insertions of $2g-2$
of these operators at genus $g$.  The spectral flow operator is constructed
entirely out of the $U(1)$ piece of the $N=2$ theory and since
the spectral flow operator can be written as
$$\sigma ={\rm exp}(i\rho/2)\qquad J=\partial \rho,$$
the twisting becomes equivalent to modifying the stress tensor
by
$$T\rightarrow T+{\partial^2 \rho \over 2},$$
where $J$ is the $U(1)$ current of $N=2$.  With this change
in the energy momentum tensor the central charge of the theory
becomes zero.
Once one does this twisting
the chiral fields which are related by spectral flow operator to the
ground states of the Ramond sector become dimension 0 and form a nice
closed ring known as the chiral ring \lvw .
Given the similarities to $N=2$ we would like to explore
analogous construction for $G_2$ and $Spin(7)$.
In the $N=2$ case the main modification
in the theory was in the $U(1)$ piece of the theory.  Therefore
also here we expect the main modifications to be in the
tri-critical Ising piece for the $G_2$ and in the Ising piece
for the $Spin(7)$ case.

Let us concentrate on the sphere.
As noted above abstractly, on the sphere one can define twisted correlation
functions by insertion of two spin fields ($\sigma_{\frac{7}{16}}$
in $G_2$ case and $\sigma_{\frac{1}{2}}$ in $Spin(7)$ case)
in NS sector:

\eqn\twist{<V_1(z_1,\bar z_1)....V_n(z_n,\bar z_n)>_{twisted}=
<\sigma(0)V_1(z_1,\bar z_1)...V_n(z_n,\bar z_n)\sigma(\infty)>_{untwisted}.}

Let us check this idea by bosonizing Ising sector. First we discuss $G_2$.
Bosonized tri-critical
Ising supercurrent and stress tensor have the form:

\eqn\bosis{\Phi=e^{\frac{3i}{\sqrt 5}\varphi},}
\eqn\bosstr{X=({\partial}\varphi)^2+\frac{1}{4\sqrt 5}{\partial}^2\varphi.}
At the same time we can write down the chiral primaries in terms of boson
$\varphi$:

\eqn\vacuum{[0]=I}
\eqn\oneten{[\frac{1}{10}]=e^{\frac{i}{\sqrt 5}{\varphi},}}
\eqn\si{[\frac{6}{10}]=e^{\frac{2i}{\sqrt 5}\varphi,}}
\eqn\sevensixt{[\frac{7}{16}]=e^{\frac{-5i}{4{\sqrt 5}}\varphi,}}
\eqn\threeeight{[\frac{3}{80}]=e^{\frac{-i}{4{\sqrt 5}}\varphi}.}
Background charge is $-2\alpha_0=-\frac{1}{2\sqrt 5}$
and one can check that central
charge is correct $c=1-24\alpha_0^2=\frac{7}{10}$.
Insertion of spin fields according to \twist\ and \sevensixt\
is equivalent to a change in background
charge $-2\alpha_0 \rightarrow -2{\tilde \alpha_0}
=-\frac{3}{\sqrt 5}$,
and thus new stress-tensor that replaces $X$ is
$X_{tw}=({\partial}\varphi)^2-\frac{3}{2\sqrt 5}{\partial}^2\varphi$ with
central charge $\tilde c_{tw}=1-24\tilde \alpha^2_0=-\frac{98}{10}$.
If we compute total central charge (we don't touch
remaining sector $T_r$
by our twist) since the central charge of $T_r$ is
equal to $21/2 -7/10=98/10$ and we have not changed it
by the twisting we get
: $c_{twist}=-98/10+98/10=0$.  This is indeed remarkable!
It is the strongest hint for the existence of a topological theory.
 Obviously, before twisting we have
a minimal model and correct vertex operators are given by above formulas
dressed by screening operators (see \ref\ff{B. Feigin and D. Fucks, Func. Anal.
i ego Priloz., 17 (1983 241.}, \ref\df
{V. Dotsenko, V. Fateev, Nucl. Phys., B240 [FS12] (1984) 312.},
\ref\f{G. Felder, Nucl. Phys., B324 (1989) 548.}); screening
charges are: $\alpha_+=\frac{5}{2\sqrt 5}, \alpha_-=-\frac{2}{\sqrt 5}$.
At the same time after twisting we get a model which is not a
minimal model and
if now correlation functions of above operators
aren't  non-zero they can't be screened. Thus, after
twisting when we calculate correlation functions
we could forget about dressing
by screening operators
and do just naive computation. This simplifies the story. Vertex operators are
the same, but their dimensions are now different. We have:
\eqn\toneten{[\frac{1}{10}] \longrightarrow [-\frac{2}{5}],}
\eqn\tsi{[\frac{6}{10}] \longrightarrow [-\frac{2}{5}],}
\eqn\threehalf{[\frac{3}{2}] \longrightarrow [0].}
Note that in particular we learn that the special states we get in the
NS sector have total dimension zero in the topological theory:
$$|{1\over 10},{2\over 5}>\rightarrow |-{2\over 5},{2\over 5}>,$$
$$|{6\over 10},{2\over 5}>\rightarrow |-{2\over 5},{2\over 5}>,$$
$$|{3\over 2} ,0         >\rightarrow |0          , 0        >.$$
which is what one would expect of topological observables.  Moreover
they do seem to form a ring under multiplication.
This can be checked explicitly for example for the untwisted moduli
of the toroidal compactification discussed in the previous
section.  Concentrating on left-movers, the states of the first
type is written as $\psi^i$ for $i=1,...,7$. Now under naive
product between the $\psi^i$ there would be poles because
of contractions, but one can see that they do not contribute
to the topological amplitude because they fail to cancel
the background charge in the topological theory.  In fact the ring
they form in this case is
$$<\psi^i \psi^j \psi^k>=f_{ijk},$$
where $f_{ijk}$ are defined by $\Phi=f_{ijk}\psi^i \psi^j \psi^k$
(note that the $6/10$ states above are nothing but
the quadratic fermion terms).

The
 expressions for the shift in the dimension of the tri-critical piece
together with the fact that we have already discussed the tri-critical
content of the generators of the chiral algebra means that
we can deduce their twisted dimension.  We find that they
all have shifted to integer dimensions, another hallmark of topological
theories:
$G - dim. 1, \Phi - dim. 0, M - dim.2$, plus we got dimension 1
bosonic operator $K$. Thus, after twisting, $G$ is a candidate for BRST
current of the topological theory and
$M$ - for antighost. To prove the last statement we need to show
that OPE's of $G$ with itself, as well as $M$ with itself don't have simple
poles (or at least do not contribute to the amplitudes)
and in addition, $G$ with $M$ have the modified stress-tensor
as a residue of simple pole.  This would need to be verified.
It should also be verified that with this sense
of topological BRST invariance the above special states
in the NS sector indeed are BRST invariant.

 It is not difficult to repeat above procedure for the case of $Spin(7)$.
Bosonized Ising stress tensor has the form:

\eqn\bosnis{\tilde X= ({\partial}\varphi)^2+
\frac{1}{4\sqrt 3}{\partial}^2\varphi}
and chiral primaries are:

\eqn\isvac{[0]=I,}
\eqn\isonehalf{[\frac{1}{2}]=e^{\frac{3i}{2\sqrt 3}\varphi},}
\eqn\isonesix{[\frac{1}{16}]=e^{\frac{3i}{4\sqrt 3}\varphi}.}
Background charge is $-2\alpha_0=-\frac{1}{2\sqrt 3}$ and screening charges
are $\alpha_+=-\frac{3}{2\sqrt 3}, \alpha_-=\frac{2}{\sqrt 3}$.
Bosonized vertex operators are given by above expressions dressed
with $n_1$ screening charges of type $\alpha_+$ and $n_2$ of the type
$\alpha_-$. Insertion of
spin fields $\sigma_{\frac{1}{2}}$ according \twist\
in the picture with $n_1=6, n_2=2$ is equivalent
to a change in background charge
$-2\alpha_0 \rightarrow -2\tilde \alpha_0=-\frac{5}{2\sqrt 3}$.\foot{
Note that in $G_2$ case we had used $n_1=0, n_2=0$ picture, which was
the minimal solution in that case.} Thus,
new stress tensor is given by $\tilde X_{tw}
=({\partial}\varphi)^2+\frac{5}{4\sqrt 3}
{\partial}^2\varphi$ with central charge $\tilde
c_{tw}=1-24{\tilde \alpha_0^2}=
-\frac{23}{2}$. If we remember that the central charge of $T_r$ was
$12-\frac{1}{2}=\frac{23}{2}$ and it has remained unchanged under
our twist we will find another remarkable coincidence: total central charge
after twist is $0$! Now we can check other properties discovered above for the
case of $G_2$. Vertex operators remain the same \isvac,
\isonehalf, \isonesix, but
now they have different dimensions:

\eqn\isver{[\frac{1}{2}] \longrightarrow [-\frac{1}{2}],}
\eqn\isverr{[\frac{1}{16}] \longrightarrow [-\frac{7}{16}].}
So, special states we had in NS sector have total dimension
zero in the topological theory:

\eqn\istot{|\frac{1}{2},\frac{1}{2}> \rightarrow |-\frac{1}{2},\frac{1}{2}>,}
\eqn\istota{|\frac{1}{16},\frac{7}{16}> \rightarrow |-\frac{7}{16},\frac{7}{16}
>.}
Also, one can use relations \genfermion\ and show that the dimensions
of fermionic operators $G$ and $\tilde M$ are shifted properly:
$dim.G=1, dim.\tilde M=2$. This means that
once again $G$ is a candidate for the BRST current and $\tilde M$ - for
antighost.

\vskip 1cm

Acknowledgements: We would like to thank P. Aspinwall,
 E. Martinec, G. Moore, N. Nekrasov and
E. Witten for valuable discussions.  We also
like to thank D.D. Joyce for discussions on his work as well as for his
comments on the manuscript.  In addition C.V. would like to thank
the Institute for Advanced Study for its hospitality and
for providing a stimulating research environment
 where this
work was done.
The research of C.V. was supported in part by the Packard fellowship
and NSF grants PHY-92-18167 and PHY-89-57162. The research of S. L. Sh. has
been supported by NSF grant PHY92-45317.

 \appendix{1}{}

Below we give the result of computations mentioned in the Section 3 and
corresponding mode expansion.

For OPE we have:

\eqn\gt{G(z)K(w) = \frac{3}{(z-w)^2} \Phi(w) + \frac{1}{z-w} {\partial}{\Phi}}

\eqn\gm{G(z)M(w) = - \frac{7}{2}\frac{1}{(z-w)^4} + \frac{1}{(z-w)^2}
 (T + 4 X)(w) + \frac{1}{z-w} {\partial}X(w)}

\eqn\ft{\Phi(z)K(w) = - \frac{3}{(z-w)^2} G(w) - \frac{3}{z-w} (M +
\frac{1}{2}{\partial}G)(w)}

\eqn\fm{\Phi(z)M(w) = \frac{9}{2}\frac{1}{(z-w)^2}K(w) - \frac{1}{z-w}
(3 :G(w)\Phi(w): - \frac{5}{2} {\partial}K(w))}

\eqn\xt{X(z)K(w) = - \frac{3}{(z-w)^2} K(w) + \frac{3}{z-w} ( :G(w)\Phi(w): -
{\partial}K(w))}

\eqn\xm{\eqalign{X(z)M(w) = &- \frac{9}{2}\frac{1}{(z-w)^3} G(w) -
\frac{1}{(z-w)^2} (5M + \frac{9}{4} {\partial}G)(w) + \cr
&\frac{1}{z-w}(4 :G(w)X(w): - \frac{7}{2} {\partial}M(w) - \frac{3}{4}
{\partial}^2G(w))}}

\eqn\tt{K(z)K(w) = - \frac{21}{(z-w)^4} + \frac{6}{(z-w)^2} (X-T)(w) +
\frac{3}{z-w}{\partial}(X-T)(w)}

\eqn\tm{\eqalign{K(z)M(w) = &- \frac{15}{(z-w)^3} \Phi(w) -
\frac{11}{2}\frac{1}{(z-w)^2} {\partial}\Phi(w)
+ \frac{3}{z-w}(:G(w)K(w): +\cr
&2:T(w)\Phi(w):)}}

\eqn\mm{\eqalign{M(z)M(w) = &- \frac{35}{(z-w)^5} + \frac{1}{(z-w)^3}
(20X-9T)(w)
+ \frac{1}{(z-w)^2}(10{\partial}X-\frac{9}{2}{\partial}T)(w)
+ \cr
&\frac{1}{z-w}(\frac{3}{2}{\partial}^2X(w) -
\frac{3}{2}{\partial}^2T(w) - 4:G(w)M(w): + 8:T(w)X(w):)}}

\eqn\TM{T(z)M(w) = - \frac{1}{2}\frac{1}{(z-w)^3}G(w) +
\frac{5}{2}\frac{1}{(z-w)^2}M(w) + \frac{1}{z-w}{\partial}M(w)}

In the case of $Spin7$ algebra looks simpler:

\eqn\pphi{\tilde X = \psi^8\Phi - X + 1/2 {\partial}\psi^8\psi^8}
\eqn\tt{\tilde M= J^8\Phi - \psi^8K - M +1/2{\partial}J^8\psi^8 -
1/2 J^0{\partial}\psi^0}

\eqn\pphipphi{\tilde X(z)\tilde X(w) = 16/(z-w)^4 +
16/(z-w)^2\tilde X(w) + 8/(z-w){\partial}\tilde X(w)}

\eqn\Tpphi{T(z)\tilde X(w) = 2/(z-w)^4 + 1/(z-w)^2(\tilde X(w) +\tilde X(z))}

\eqn\Gpphi{G(z)\tilde X(w) = 1/2(z-w)^2G(w) + 1/(z-w)\tilde M(w)}

\eqn\Gm{G(z)\tilde M(w)=\frac{4}{(z-w)^4}-\frac{1}{(z-w)^2}(T(w)-4\tilde X(w))
+\frac{1}{z-w}{\partial}\tilde X(w)}

\eqn\pphitt{\eqalign{ \tilde X(z)\tilde M(w) &= - 15/2(z-w)^3G(w) -
1/(z-w)^2(15/4{\partial}G(w) -\cr
&8\tilde M(w)) + 1/(z-w)(11/2{\partial}\tilde M(w) -
5/4{\partial}^2G(w) - \cr
&6 :G(w)\tilde X(w):)}}

\eqn\mm{\eqalign{\tilde M(z)\tilde M(w) =& - 64/(z-w)^5 - 1/(z-w)^3(15T(w) +
32\tilde X(w))
-\cr
&1/(z-w)^2(15/2{\partial}T(w) +16{\partial}\tilde X(w)) -
1/(z-w)(5/2{\partial}^2\tilde X(w) + \cr
&5/2{\partial}^2T(w) + 12:T(w)\tilde X(w):- 6:G(w)\tilde M(w):)}}

Now, if we use mode expansion for our generators $B(z)=B_nz^{-n-\Delta}$,
where $\Delta$ is a dimension of operator $B$, we have
(we use the normal ordering prescription $:AB:_n=\sum_{p<-\Delta_A-1}A_pB_{n-p}
+(-1)^{AB}\sum_{p>-\Delta_A}B_{n-p}A_p$) :

\eqn\conj{\Phi_n^{+}=-\Phi_{-n}, \quad \quad
K_n^{+}=-K_{-n}, \quad \quad M_n^{+}=\frac{1}{2}G_{-n}-M_{-n}}
\eqn\mgg{\{G_n,G_m\} = \frac{7}{2}(n^2-\frac{1}{4})\delta_{n+m,0}+2L_{n+m}}
\eqn\mTT{[L_n,L_m]=\frac{21}{24}(n^3-n)\delta_{n+m,0}+(n-m)L_{n+m}}
\eqn\mTg{[L_n,G_m]=(\frac{1}{2}n-m)G_{n+m}}
\eqn\phiph{\{\Phi_n,\Phi_m\}=-\frac{7}{2}(n^2-\frac{1}{4})\delta_{n+m,0}+
6X_{n+m}}
\eqn\Xphi{[X_n,\Phi_m]=-5(\frac{1}{2}n-m)\Phi_{n+m}}
\eqn\XX{[X_n,X_m]=\frac{35}{24}(n^3-n)\delta_{n+m,0}-5(n-m)X_{n+m}}
\eqn\TX{[L_n,X_m]=-\frac{7}{24}(n^3-n)\delta_{n+m,0}+(n-m)X_{n+m}}
\eqn\gphi{\{G_n,\Phi_m\}=K_{n+m}}
\eqn\gt{[G_n,K_m]=(2n-m)\Phi_{n+m}}
\eqn\gX{[G_n,X_m]=-\frac{1}{2}(n+\frac{1}{2})G_{n+m}+M_{n+m}}
\eqn\gM{\{G_n,M_m\}=-\frac{7}{12}(n^2-\frac{1}{4})(n-\frac{3}{2})
\delta_{n+m,0}+
(n+\frac{1}{2})L_{n+m}+(3n-m)X_{n+m}}
\eqn\phit{[\Phi_n,K_m]=\frac{3}{2}(m-n+\frac{1}{2})G_{n+m}-3M_{n+m}}
\eqn\phiM{\{\Phi_n,M_m\}=(2n-\frac{5}{2}m-\frac{11}{4})K_{n+m}-3:G\Phi:_{n+m}}
\eqn\xt{[X_n,K_m]=3(m+1)K_{n+m}+3:G\Phi:_{n+m}}
\eqn\XM{\eqalign{[X_n,M_m]=[\frac{9}{4}(n+1)&
(m+\frac{3}{2})-\frac{3}{4}(n+m+\frac{3}{2})
(n+m+\frac{5}{2})]G_{n+m}-[5(n+1)-\cr
&\frac{7}{2}(n+m+\frac{5}{2})]M_{n+m}+4:GX:_{n+m}}}
\eqn\tt{[K_n,K_m]=-\frac{21}{6}(n^3-n)\delta_{n+m,0}+3(n-m)(X_{n+m}-L_{n+m})}
\eqn\tM{[K_n,M_m]=[\frac{11}{2}(n+1)(n+m+\frac{3}{2})-\frac{15}{2}(n+1)n]
\Phi_{n+m}+3:GK:_{n+m}-6:L\Phi:_{n+m}}
\eqn\MM{\eqalign{\{M_n,&M_m\}=-\frac{35}{24}(n^2-\frac{1}{4})(n^2-\frac{9}{4})
\delta_{n+m,0}+[\frac{3}{2}(n+m+2)(n+m+3)-\cr
&10(n+\frac{3}{2})(m+\frac{3}{2})]X_{n+m}+[\frac{9}{2}(n+
\frac{3}{2})(m+\frac{3}{2})-\frac{3}{2}(n+m+2)(n+m+3)]L_{n+m}-\cr
&4:GM:_{n+m}+8:LX:_{n+m}}}

We have similar equations for the case of $Spin(7)$.
Operator $\tilde M$ again has nonstandard conjugation property

\eqn\conjj{\tilde M^{+}_n=-\frac{1}{2}G_{-n}-\tilde M_{-n},}
and commutation relations are given by:

\eqn\mxx{[\tilde X_n,\tilde X_m]=\frac{16}{6}(n^3-n)\delta_{n+m,0}+8(n-m)\tilde
X_{n+m},}
\eqn\lxx{[L_n,\tilde X_m]=\frac{1}{3}(n^3-n)\delta_{n+m,0}+(n-m)\tilde
X_{n+m},}
\eqn\gxx{[G_n,\tilde X_m]=\frac{1}{2}(n+\frac{1}{2})G_{n+m}+\tilde M_{n+m},}
\eqn\gmm{\{G_n,\tilde M_m\}=\frac{2}{3}(n^2-\frac{1}{4})(n-\frac{3}{2})
\delta_{n+m,0}-(n+\frac{1}{2})L_{n+m}+(3n-m)\tilde X_{n+m},}
\eqn\xmm{\eqalign{[\tilde X_n,\tilde M_m]=&[\frac{15}{4}(n+1)(m+\frac{3}{2})-
\frac{5}{4}(n+m+\frac{3}{2})(n+m+\frac{5}{2})]G_{n+m}-\cr
&[8(n+1)+\frac{11}{2}(n+m+\frac{5}{2})\tilde M_{n+m}-6:G\tilde X:_{n+m},}}
\eqn\mmm{\eqalign{\{\tilde M_n,\tilde M_m\}=&-\frac{8}{3}(n^2-\frac{9}{4})
(n^2-\frac{1}{4})
+[\frac{15}{2}(n+\frac{3}{2})(m+\frac{3}{2})-\cr
&\frac{5}{2}(n+m+2)(n+m+3)]L_{n+m}+\cr
&[16(n+\frac{3}{2})(m+\frac{3}{2})-\cr
&\frac{5}{2}(n+m+2)(n+m+3)]
\tilde X_{n+m}+12:L\tilde X:_{n+m}-\cr
&6:G\tilde M:_{n+m}.}}

\appendix{2}{}

First, let us note that from commutation relations, given in Appendix 1, and
\conj, \conjj\ we could derive
following identities:

\eqn\anih{\eqalign{&|M_{-1/2}(0,0)|^2=(0,0)^*M^{+}_{1/2}M_{-1/2}(0,0)=0,\cr
&|M_{-3/2}(0,0)|^2=(0,0)^*M^{+}_{3/2}M_{-3/2}(0,0)=0,}}
\eqn\annih{\eqalign{&|\tilde M_{-1/2}(0,0)|^2=
(0,0)^*{\tilde M^{+}_{1/2}}{\tilde M_{-1/2}}(0,0)=0,\cr
&|\tilde M_{-3/2}(0,0)|^2=
(0,0)^*{\tilde M^{+}_{3/2}}{\tilde M_{-3/2}}(0,0)=0.}}
These identities are necessary because as we had already seen operators $M$ and
$\tilde M$ have nonstandard conjugation properties and in principle
$M_{-1/2}, M_{-3/2}, \tilde M_{-1/2}, \tilde M_{-3/2}$ might not annihilate
the vacuum. But we see that they do.

 Finally, we will show the validity of \gener, \generfer, \gengen\
and \genfermion. First two identities in \gener\ and \gengen\
are obvious and to
derive the last one in \gener\ and \generfer\
(\genfermion) we have to apply
zero mode $T_0^I=-\frac{1}{5}X_0$ ($\tilde T_0^I=\frac{1}{8}{\tilde X_0}$)
to the left hand side:

\eqn\der{\eqalign{-\frac{1}{5}X_0G_{-3/2}|0,0>=
\frac{1}{10}&G_{-3/2}|0,0>-\frac{1}{5}M_{-3/2}|0,0>=\cr
&
=\frac{1}{10}G_{-3/2}|0,0>,}}
\eqn\derr{\eqalign{\frac{1}{8}{\tilde X_0}G_{-3/2}|0,0>=\frac{1}{16}&G_{-3/2}
|0,0>+\frac{1}{8}M_{-3/2}|0,0>=\cr
&
=\frac{1}{16}G_{-3/2}|0,0>.}}
Here we used \anih\ and \annih. Another useful relation is \gX\
(for $Spin(7)$ - \gxx) which leads to:

\eqn\last{\eqalign{M_{-5/2}|0,0>=-X_{-1}G_{-3/2}|0,0>-&\frac{1}{2}
L_{-1}G_{-3/2}|0,0>=\cr
&
=(-\frac{7}{10}X_{-1}-\frac{1}{2}L^r_{-1})G_{-3/2}|0,0>,}}
\eqn\lastt{\eqalign{\tilde M_{-5/2}|0,0>=
-\tilde X_{-1}G_{-3/2}|0,0>+&\frac{1}{2}L_{-1}G_{-3/2}|0,0>=\cr
&
=(-\frac{7}{16}\tilde X_{-1}+\frac{1}{2}L^r_{-1})G_{-3/2}|0,0>.}}
For $K_{-2}|0,0>$ we simply use:
\eqn\k{K_{-2}|0,0>=\Phi_{-1/2}G_{-3/2}|0,0>.}
{}From \der\ and \derr\ it follows that $G_{-3/2}|0,0>$ is a
linear combination of $|\frac{1}{10},\frac{14}{10}>$
($|\frac{1}{16},\frac{23}{16}>$) and $|0,\frac{3}{2}>$,
but latter can be excluded because there is no half-integer chiral
spin $\frac{3}{2}$ operator in the $T^r$ sector of our theory.
Thus, $G_{-3/2}|0,0>=|\frac{1}{10},\frac{14}{10}>$ and
relations \gener, \generfer, \gengen\ and \genfermion\ are
consequences of above computations.

\listrefs

\end